\documentclass[preprint,review,12pt]{elsarticle}

\usepackage{amssymb}
\usepackage{amsmath}
\usepackage{graphicx}
\usepackage[retainorgcmds]{IEEEtrantools}
\usepackage[colorlinks]{hyperref}
\hypersetup{
  citecolor = {blue}
}
\usepackage{siunitx}
\usepackage{pdflscape}

\usepackage{enumitem}
\usepackage[normalem]{ulem}
\useunder{\uline}{\ul}{}
\usepackage{setspace} 
\usepackage[utf8]{inputenc}
\usepackage[linesnumbered, ruled, vlined]{algorithm2e}
\usepackage{rotating}
\usepackage{nicefrac}
\usepackage{xspace}
\usepackage{float}
\usepackage{caption}
\usepackage{subcaption}
\usepackage{rotating, lscape, longtable, tabu, booktabs, siunitx, multirow}
\usepackage[capitalise, noabbrev]{cleveref}
\usepackage{verbatim}
\usepackage{mathtools}
\usepackage{natbib}
\usepackage{array}
\usepackage[T1]{fontenc}
\usepackage[utf8]{inputenc}
\usepackage{fancyhdr} 
\usepackage{lipsum}
\usepackage{xcolor}
\usepackage{lineno}

\usepackage{tkz-graph}

\usetikzlibrary{arrows,arrows.meta,shapes,decorations.pathmorphing, decorations.markings, positioning}
\let\svtikzpicture\tikzpicture
\def\tikzpicture{\noindent\svtikzpicture}

\newcommand{\uset}[1]{\ifmmode\left\{\,#1\,\right\}\else\{\,#1\,\}\fi}
\newcommand{\ulst}[1]{\ifmmode\left[\,#1\,\right]\else[\,#1\,]\fi}
\newcommand{\upar}[1]{\ifmmode\left(\,#1\,\right)\else(\,#1\,)\fi}
\newcommand{\uioc}[1]{\ifmmode\left(\,#1\,\right]\else(\,#1\,]\fi}
\newcommand{\uico}[1]{\ifmmode\left[\,#1\,\right)\else[\,#1\,)\fi}

\journal{Advances in Water Resources}

\begin{document}

\begin{frontmatter}

\title{Pore-scale imaging of hydrogen and methane storage in fractured aquifer rock: the impact of gas type on relative permeability}

\author[1,2]{Sojwal Manoorkar \corref{cor1}}
\ead{sojwal.manoorkar@Ugent.be/sojwal.m@gmail.com}
\author[1,2]{G{\"u}lce Kalyoncu Pakkaner}
\author[1,2]{Hamdi Omar}
\author[1,2]{Soetkin Barbaix}
\author[3]{Dominique Ceursters}
\author[3]{Maxime Latinis}
\author[3]{Stefanie Van Offenwert}
\author[1,2]{Tom Bultreys}

\address[1]{Department of Geology, Ghent University, Ghent, Belgium}
\address[2]{Centre for X-ray Tomography (UGCT), Ghent University, Ghent, Belgium}
\address[3]{Fluxys, Belgium}

\cortext[cor1]{I am corresponding author}

\begin{abstract}

Underground hydrogen storage in saline aquifers is a potential solution for seasonal renewable energy storage. Among potential storage sites, facilities used for underground natural gas storage have advantages, including well-characterized cyclical injection-withdrawal behavior and partially reusable infrastructure. However, the differences between hydrogen-brine and natural gas-brine flow, particularly through fractures in the reservoir and the sealing caprock, remain unclear due to the complexity of two-phase flow. Therefore, we investigate fracture relative permeability for hydrogen versus methane (natural gas) and nitrogen (commonly used in laboratories).  Steady-state relative permeability experiments were conducted at 10 MPa on fractured carbonate rock from the Loenhout natural gas storage in Belgium, where gas flows through {\textmu}m-to-mm scale fractures. Our results reveal that the hydrogen exhibits similar relative permeability curves to methane, but both are significantly lower than those measured for nitrogen. This implies that nitrogen cannot reliably serve as a proxy for hydrogen at typical reservoir pressures. The low relative permeabilities for hydrogen and methane indicate strong fluid phase interference, which traditional relative permeability models fail to capture. This is supported by our observation of periodic pressure fluctuations associated with intermittent fluid connectivity for hydrogen and methane. In conclusion, our findings suggest that the fundamental flow properties of fractured rocks are complex but relatively similar for hydrogen and natural gas. This is an important insight for predictive modeling of the conversion of Loenhout and similar natural gas storage facilities, which is crucial to evaluate their hydrogen storage efficiency and integrity.

\end{abstract}


\end{frontmatter}

\section{Introduction} \label{sec:intro}

The hydrogen economy is expected to play a critical role in the global transition to cleaner energy and achieving net-zero carbon emissions. A key technology supporting large-scale energy storage is Underground Hydrogen Storage (UHS). UHS enables the storage of TWh-scale energy by injecting hydrogen into depleted gas reservoirs, saline aquifers, and existing gas storage facilities. The stored (green) hydrogen can be produced by electrolysis during periods of surplus electricity generation \cite{Tarkowski2019, Tarhan2021, Zivar2021, Thiyagarajan2022, Al-Shafi2023, Perera2023, Ali2025}. As renewable energy sources like wind and solar continue to expand \cite{Acar2014}, UHS provides a potential solution to balance renewable energy supply and demand, enhancing energy security, and helping to decarbonize energy systems. Within the broader hydrogen economy, emerging subsurface strategies such as in situ hydrogen generation from hydrocarbons \cite{Hassanpouryouzband2025} and the conceptual framework of engineered `hydrogen farming' \cite{Hassanpouryouzband2024} further reinforces hydrogen's role as a critical energy vector for net-zero transitions.

Converting pre-existing natural gas storage sites to hydrogen storage potentially offers advantages to using depleted gas fields or saline aquifers that have not been previously deployed for storage purposes. This includes leveraging expensive site characterization campaigns, tried-and-tested reservoir and caprock behaviour under cyclical injection-retraction operations, and existing connections to gas treatment and pipeline infrastructure. Among operational and planned natural gas storage sites in Europe, 86\% are depleted gas fields and 14\% are aquifers, together capable of providing up to 750 TWh of hydrogen storage \cite{HyUSPre2022}. Conversion of these storage resources would be sufficient for a mid-range scenario of 2,500 TWh of annual hydrogen demand by 2050, at a 30\% storage capacity. In this paper, we focus on a notable example of such a natural gas storage site under investigation for hydrogen conversion: the Loenhout natural gas storage facility in northern Belgium. Loenhout is uniquely well-connected in Western Europe as it is part of a pipeline network linking Belgium, France, The Netherlands, Germany, Luxemburg and the United Kingdom, including international shipping terminals in the North Sea ports of Zeebrugge and Dunkirk. If converted succesfully, it would provide 2.2 TWh of storage capacity. The site comprises a partially karstified limestone aquifer from Visean age, characterized by a fracture-dominated flow system with low primary porosity \cite{Dreesen1987, Lagrou2012, Vandenberghe2014}. Its conversion is currently under investigation in the BE-HyStore project funded by the Belgian federal Energy Transition Fund and the site operator Fluxys Belgium. 

Repurposing an aquifer natural gas storage site for large-scale hydrogen injection requires evaluating whether the flow of the stored gas in the pores or fractures of the storage reservoir would be affected, as this may impact injectivity and operational efficiency \citep{Heinemann2021a, Lysyy2021}. While significant progress has been made in understanding hydrogen flow in porous media, substantial knowledge gaps remain regarding its behavior in fractured rock. This is not only important for fractured storage reservoirs such as in Loenhout, but also to assess leakage through caprock if fractures are present or generated during storage \cite{Aftab2022, Aslannezhad2023}. A primary uncertainty in the current scientific understanding lies in hydrogen's relative permeability compared to methane’s. Relative permeability is a key parameter for upscaled multiphase flow modeling, which quantifies how the flow of gas is impacted by the presence of (saline) groundwater in the reservoir, and vice versa \citep{Blunt2017}. 

Over the last two decades, several studies have shown that the relative permeability behaviour of fractured media differs significantly from that of porous materials due to its dependence on complex fluid dynamics \citep{Phillips2020}. Nevertheless, it is currently unclear how such complexities affect fracture relative permeabilities for hydrogen compared to other gases, particularly at elevated pressures and temperatures typical for UHS. Experiments with air-water in transparent fractures have revealed strong phase interference and reduced relative permeability at intermediate saturations in rough fractures \citep{Pruess1990, Persoff1995}. Relative permeability in such systems is highly sensitive to fracture surface roughness and aperture heterogeneity, influencing the meandering or twisting of flow paths \citep{Chen2006, Huo2016, Nicholl2000, Yang2013, Ye2015}. Capillary-dominated flow, especially in low viscosity-ratio fluid systems such as hydrogen-brine and methane-brine, is characterized by fluid pathways that are intermittently connected, caused by snap-off events and Haines jumps \citep{Phillips2023}. State-of-the-art fracture relative permeability models aim to capture the greater phase interference, the absence of percolating pathways, and altered flow patterns compared to porous media \citep{Hatami2022, Watanabe2015, Wang2018, Gong2021, Eliebid2024}. The current need for  understanding how the findings above translate to UHS is underscored by a lack of experimental studies on fractured rock. This is despite significant recent efforts on the corresponding behaviour of porous reservoir rocks \citep{Yekta2018, Lysyy2022, Boon2022, Jangda2023, Higgs2024, Higgs2023, Bo2023, Rezaei2022}, which are difficult to translate directly to fracture systems. Comparative studies of two-phase flow behavior of relevant fractured samples under consistent reservoir conditions are thus critical to address these uncertainties. 

Assessing hydrogen flow in fractured aquifers such as Loenhout is essential to evaluate its feasibility as a storage medium in comparison to methane. Therefore, our study presents the first direct measurements of drainage relative permeability for hydrogen-brine, methane-brine, and nitrogen-brine systems in fractured limestone rock at subsurface pressure conditions of 10 MPa. The investigated rock used was collected from a depth of \SI{1400}{\meter} at the Loenhout aquifer. Methane was selected because it is a good model for the stored gas in current natural gas storage aquifers \citep{Hashemi2021}, and it may serve effectively as a cushion gas. Nitrogen was selected because of its common use in laboratory experiments, and because it is a potential cushion gas in new storage sites \citep{Zamehrian2022}. Using X-ray computed tomography (XCT), we captured pore-scale fluid distribution within fractures during steady-state drainage. The recorded pressure fluctuations are analyzed to identify frequency of flow intermittency. By comparing the relative permeability and flow behavior of hydrogen, methane, and nitrogen, this study contributes to a deeper understanding of gas-brine interactions in fractured systems and offers critical data to improve the accuracy of field-scale UHS models.

\section{Materials and methods} \label{sec:methods}

\subsection{Rocks and fluids}

A partially karstified limestone from the Viséan (Dinantian) age was drilled from well DZH24 at the underground natural gas storage facility in Loenhout, Belgium. This limestone belongs to the Carboniferous group in the Campine-Brabant Basin of northern Belgium. The rock matrix exhibits very low porosity ($< $0.01) and permeability ($< $1 mD), indicating that fluid flow predominantly occurs through its fracture network. Previous reservoir characterization has shown that joints and partially open veins contribute to the overall permeability \cite{VanderVoet2020}. Since naturally fractured samples were unavailable for this study, artificially fractured samples were used as the best alternative for conducting our experiments.

\begin{figure}[H] 
\centering
\begin{subfigure}[b]{0.38\textwidth}
 \centering
 \includegraphics[width=\textwidth]{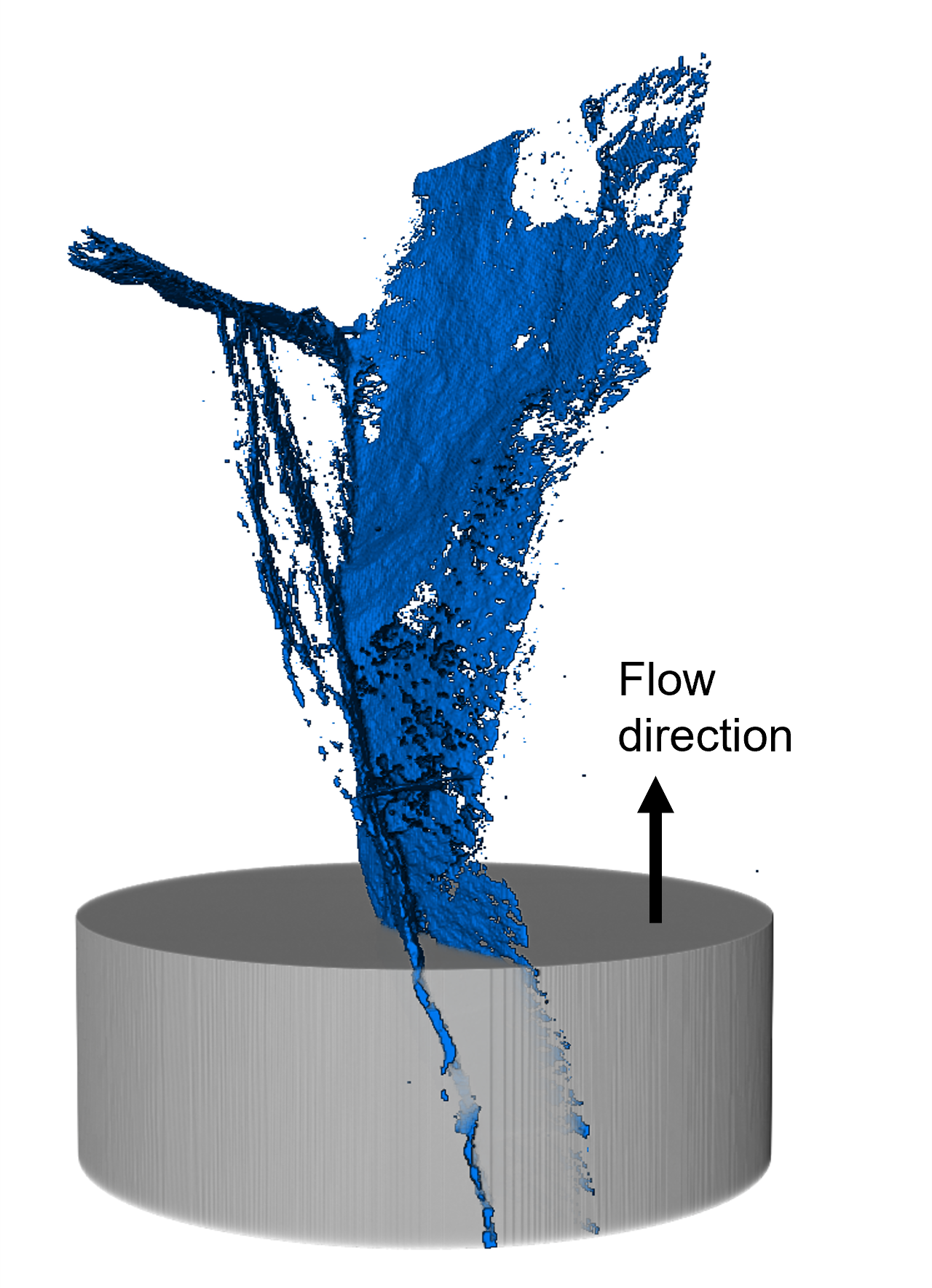}
  \caption{}
 \label{Dry3DScan}
\end{subfigure} \hspace{5mm}
\begin{subfigure}{0.47\textwidth}
 \centering
 \includegraphics[width=\textwidth]{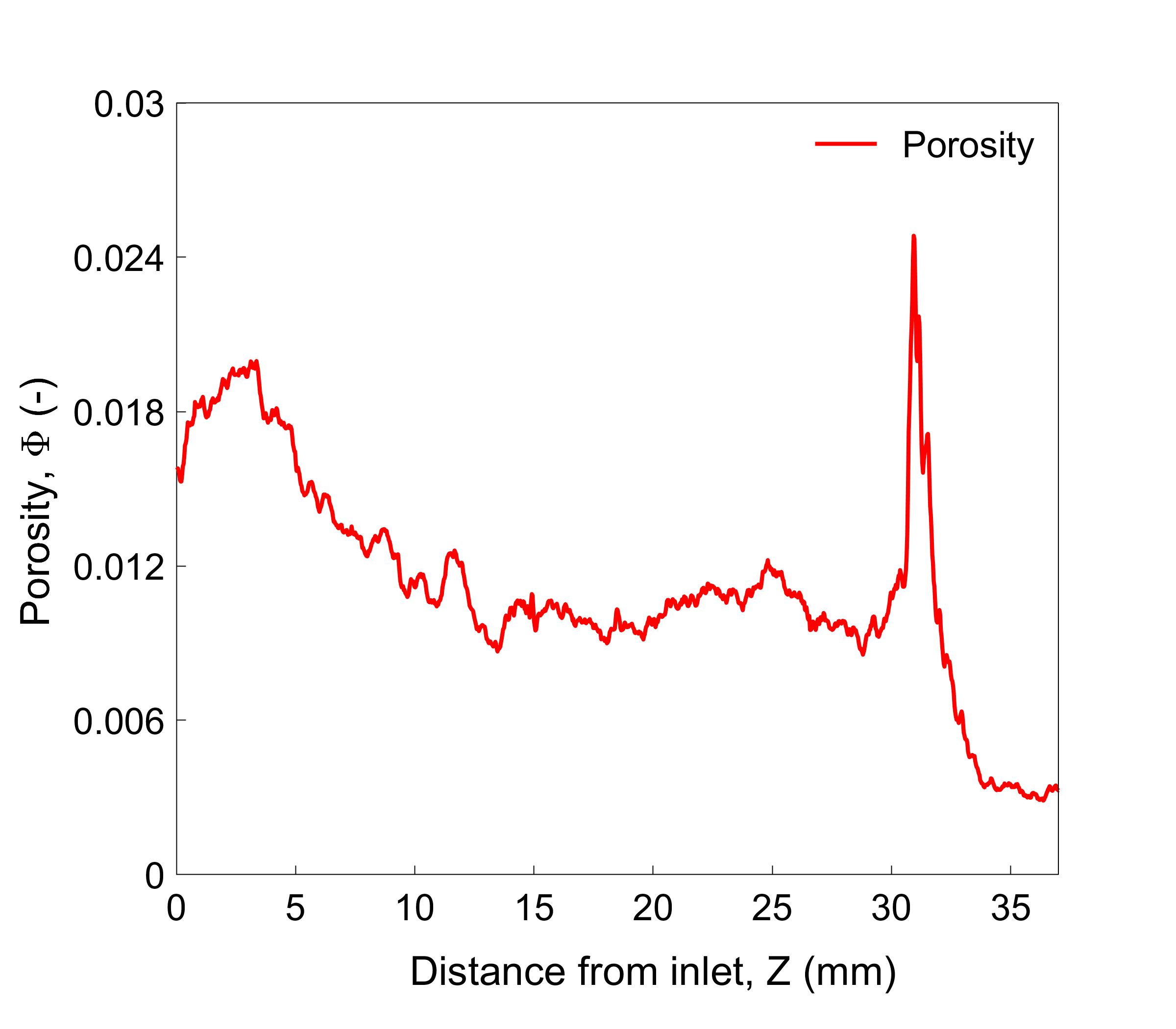}
  \caption{}
 \label{DryScan_Porosity}
\end{subfigure}
\begin{subfigure}{0.47\textwidth}
\centering  
\includegraphics[width=\textwidth]{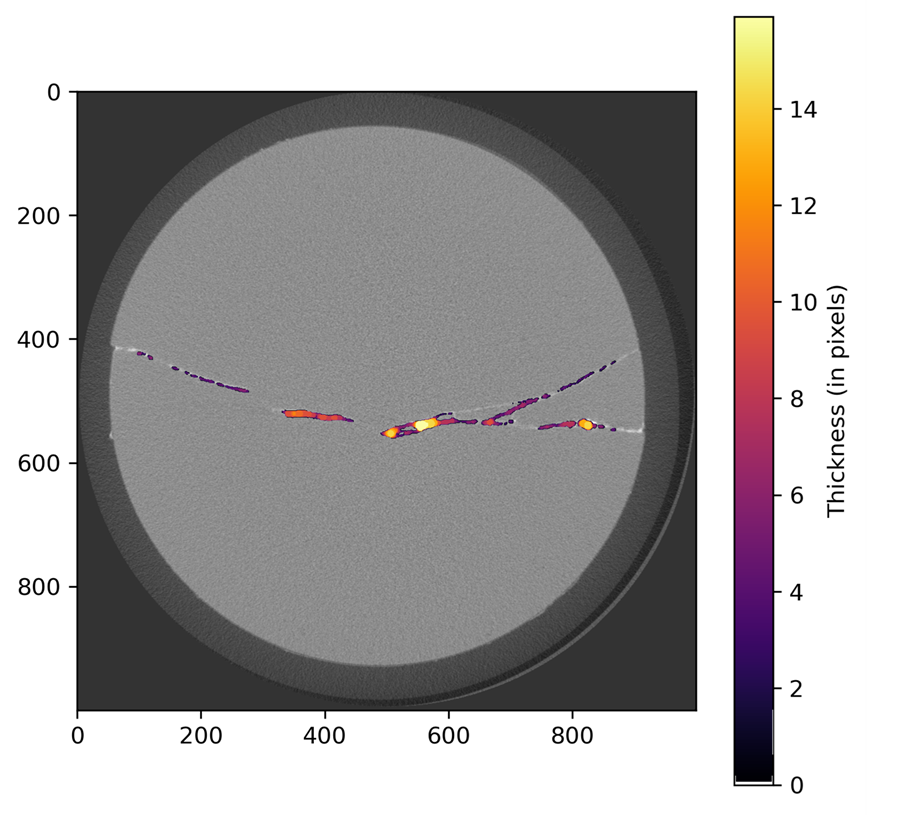} 
   \caption{}
\label{Thickness_885}
\end{subfigure} 
\begin{subfigure}{0.47\textwidth}
 \centering
 \includegraphics[width=\textwidth]{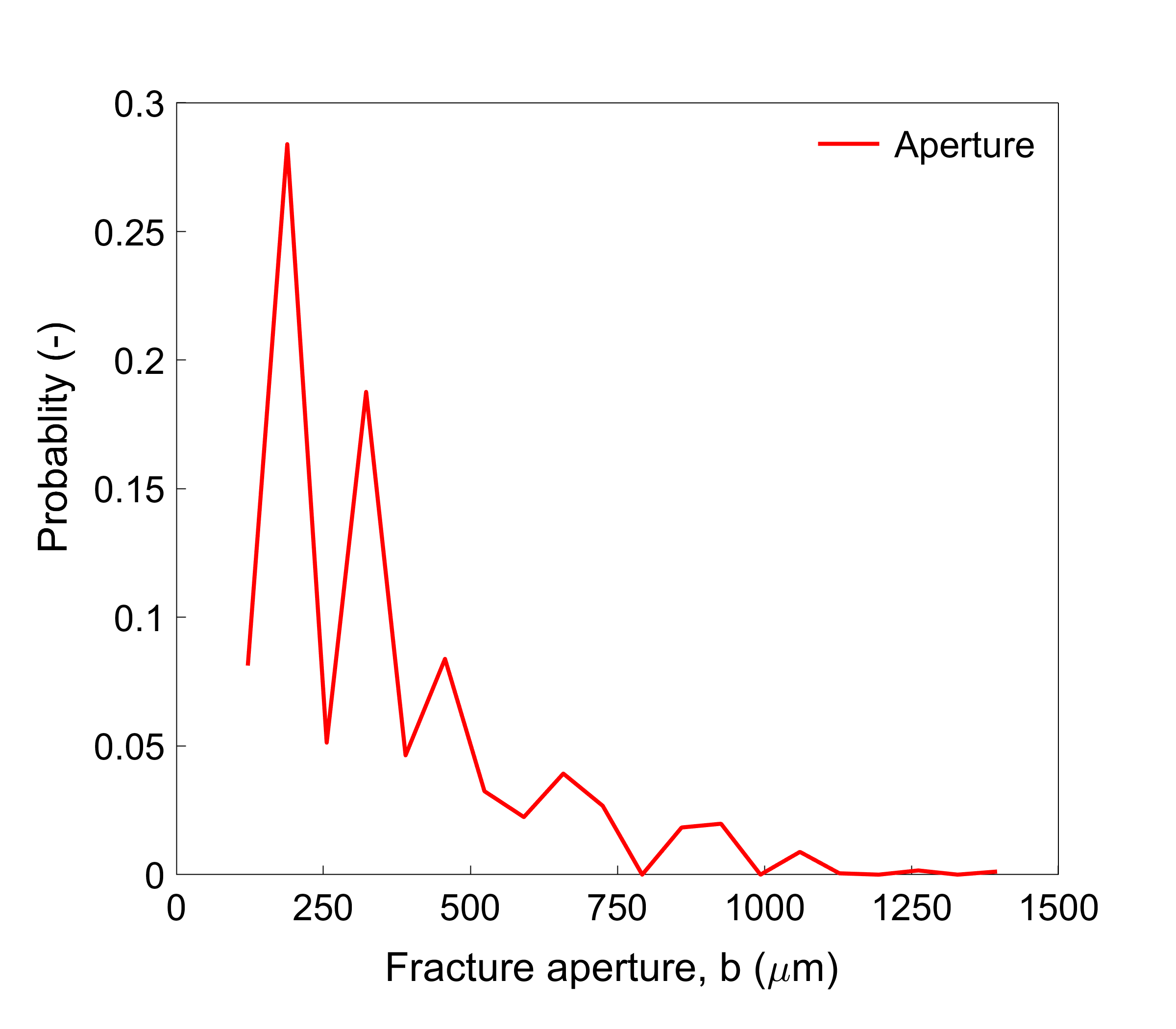} 
  \caption{}
\label{Aperture}
\end{subfigure}
\caption{(a) 3D rendering of fracture network of the core. (b) Porosity along the length of the core $<\Phi> = 0.011$.  (c) 2D local thickness map of a horizontal slice of dry scan. (d) Aperture (local thickness) distribution with $<b>$ = \SI{350.}{\micro\metre}.}
\label{Aperture_distribution}
\end{figure}

A cylindrical rock core, measuring \SI{25}{\milli\metre} in diameter and \SI{45}{\milli\metre} in length, was drilled for the experiments in this study. The core was wrapped in Teflon tape and fractured using the Brazilian tensile stress method at Ghent University’s concrete lab. This technique was selected to generate a fracture network that closely mimics natural conditions, with variable fracture width and surface roughness as shown in Figure \ref{Dry3DScan}. The dry fracture image and the porosity distribution along the core are shown in Figure \ref{DryScan_Porosity}, with the average measured porosity being 0.011. Flow through rough fractures is governed by variable aperture and relative roughness, characterized by the roughness index, $\lambda_b = \frac{<b>^\sigma}{<b>}$, where $<b>^\sigma$ represents the standard deviation of the aperture distribution, and $<b>$ is the mean aperture. Aperture was obtained from the segmented dry scan using the `PoreSpy' Python toolkit \cite{Gostick2019}, which calculates local thickness as shown in Figure \ref{Thickness_885}. The aperture distribution is shown in Figure \ref{Aperture}, with a mean aperture $<b>$ of \SI{350}{\micro\metre}. The relative roughness index was calculated to be 0.61.

The wetting phase was prepared by dissolving 25 wt$\%$ potassium iodide (Sigma Aldrich) in deionized water, which enhanced X-ray contrast for imaging purposes. The brine was selected to provide optimal contrast for X-ray imaging. The composition of brine defers from reservoir brine but sample is strongly water-wet which is also the case with realistic reservoir brine \cite{Ringrose2020, Worden2024}. The salinity of 25 wt$\%$ KI solution in terms of mole fraction is 0.0349 (34900 ppm) which falls within the typical range reported for saline aquifers in the literature \cite{Aftab2022}. The brine was not pre-equilibrated with gas prior to injection, due to limitation of our experiment to do this for large volumes of fluids needed for co-injection. This means that the co-injected fluids equilibrate as they flow through the core. While this is relevant to near wellbore regions, the measurements may not fully representative of hydrogen-brine flow deeper in the reservoir. For example, gas dissolution into the brine could potentially lead to reduced gas connectivity close to sample inlet. However, given hydrogen’s low solubility in water and employed flow rates, the flow behavior is expected to be dominated by hydrodynamic processes, with limited dissolution.

As non-wetting phases, high-purity ($>$99.99$\%$) H$_2$, CH$_4$ and N$_2$ gases (supplied by Air Liquide) were used. The thermophysical properties of these gases are listed in Table \ref{thermophysical+properties}. We maintained the same flow rate instead of matching the capillary number, as the reservoir properties are fixed. Given that nitrogen and methane are used as cushion gases, it is more practical to keep the injection rate consistent rather than focusing on the capillary number. Notably, the capillary number for all flow rates and gases is less than 10$^{-5}$, indicating that the flow is capillary-dominated and exhibits similar flow regimes. It should be noted that the experimental temperature does not match the reservoir temperature due to technical limitations. However, both hydrogen and methane remain in the gaseous state, and nitrogen remains in the supercritical state under the conditions used in this study, as they would at full reservoir conditions. The interfacial tension (IFT) of the H$_2$-water system and the viscosity of H$_2$ gas at ambient temperature are only marginally higher by less than 6\% \cite{Hosseini2022, Chow2018} and 5\% \cite{H2tool}, respectively than at reservoir temperature. Therefore, the impact of temperature on IFT and viscosity is minimal. This negligible influence is also supported by similar core-flood experiments conducted under comparable conditions \citep{Bijay2025}.

\begin{table}
 \caption{Thermophysical properties of fluids at temperature 295K and pressure 100 bar}
\centering
 \begin{tabular}{c  c  c c  c }
\hline
 Fluid  & Density       & Viscosity    & Viscosity ratio  & Interfacial tension  \\
           &  kg m$^{-3}$   &   Pa s       &        $M$       &     N m$^{-1}$  \\
 \hline
   H$_2$ \cite{Hosseini2022} & 7 & 9.8 x 10$^{-6}$ &0.01 & 0.072  \\
   CH$_4$ \cite{Bjorkvik2023} & 61 & 1.44 x 10$^{-5}$ & 0.016 & 0.05  \\
   N$_2$ \cite{Chow2016}  & 112 & 1.5 x 10$^{-4}$ & 0.16 & 0.067  \\
  \hline
\label{thermophysical+properties}
 \end{tabular}
\end{table}

\subsection{Experimental setup and procedure}

 \begin{figure}
 \centering
 \noindent\includegraphics[width=\textwidth]{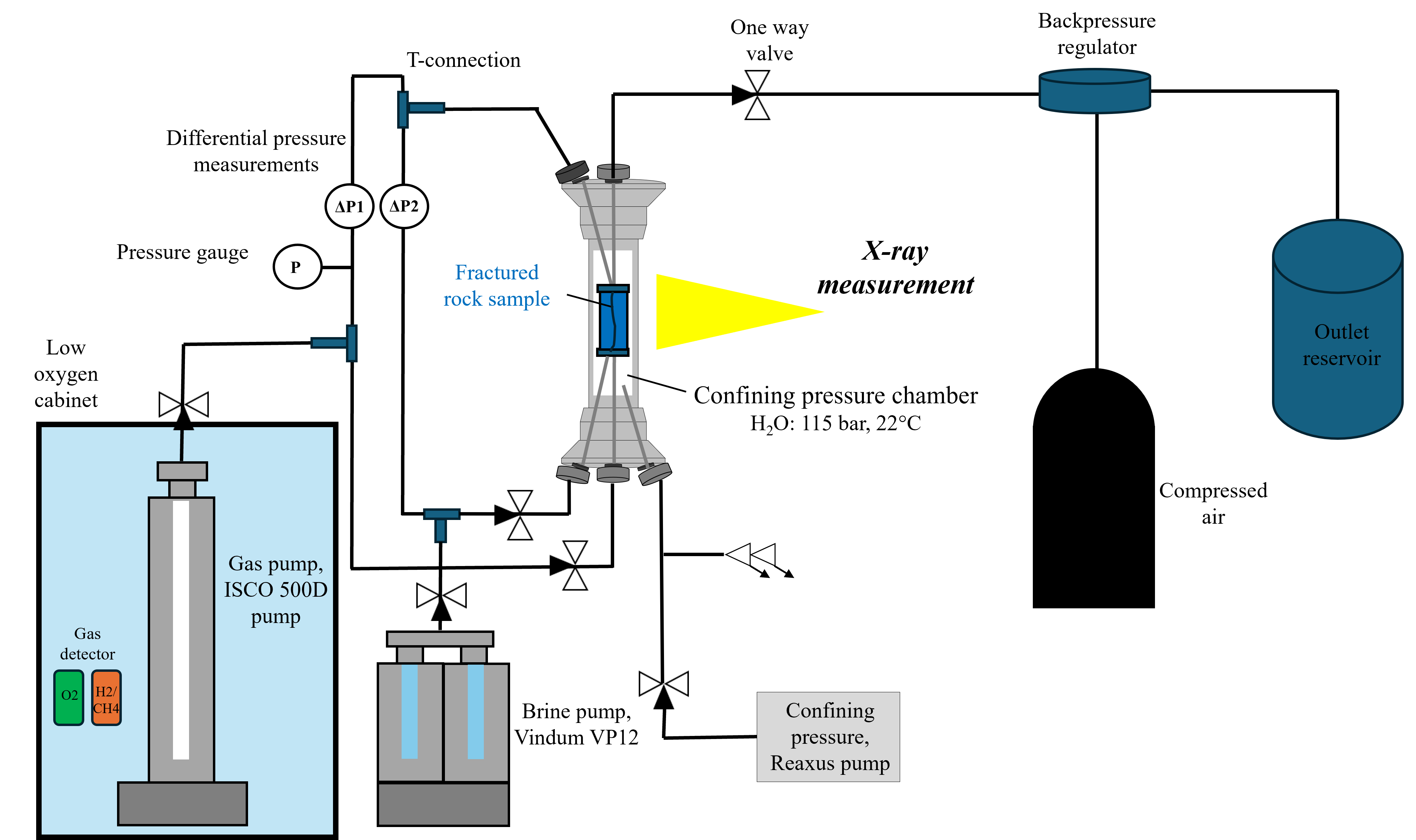}
 \caption{Experimental set-up for two-phase hydrogen/methane/nitrogen-brine core floods performed with X-ray micro computed tomography scanner.}
\label{setup}
 \end{figure}

The experimental setup is shown in figure \ref{setup}. Brine and gas are co-injected at increasing gas fractional flows to measure the drainage relative permeability, while keeping the constant total flow rate of 0.1 ml min$^{-1}$. Brine (wetting phase) is delivered using dual syringe Brine pump (Vindum Engineering VP12K). Gas (nonwetting phase) is injected using a syringe Gas pump (ISCO 500D pump). To minimize capillary end effects, \SI{2}{\milli\metre} glass bead packs were placed at both the inlet and outlet of the core as fluid distributors. Due to highly flammable nature of hydrogen and methane, the gas pump was stored inside a low-oxygen (< 4 $\%$) safety cabinet. The  detailed experimental procedure is as follows:

\begin{enumerate}
\item {Prior to experimental measurements, the fractured sample was dried in an oven overnight at 40$^{\circ}$C. The core was then wrapped in aluminum foil to prevent leakage of hydrogen gas through rubber sleeve.}
\item{A dry cylindrical fractured rock sample was placed inside a Viton rubber sleeve and loaded in the flow cell.}
\item{The confining pressure of 1.5 MPa was applied around the sample with deionized water using  a high pressure reciprocal Confining pump (Teledyne ISCO Reaxus, USA) to secure the sample in position.}
\item{A dry scan of entire length of the sample was obtained using X-ray CT scanner. }
\item{25 wt$\%$  potassium iodide (KI) brine solution was injected through the sample at volumetric flow rate of 5 ml min$^{-1}$ using Brine pump to saturate the fracture.}
\item{The flow system was  pressurized with KI brine (25 wt$\%$) upto 10 MPa while maintaining 15 bar difference between inlet and confining fluid pressures}
\item{Absolute permeability: We report liquid permeability as a reference for absolute permeability, rather than gas-phase values, to avoid the influence of gas slippage. Gas permeability can vary with gas type due to the Klinkenberg effect, leading to overestimation. As shown by \citet{Klinkenberg1941} and \citet{Letham2016}, liquid permeability provides a more accurate estimate of the absolute  permeability. The core was flooded with brine at several constant flow rates ranging from 0.1 ml min$^{-1}$ to 1 ml min$^{-1}$. Absolute permeability was then determined using Darcy's law as described in equation \ref{permeability}. The pressure gradient was monitored at steady-state using a Keller-33PDX pressure transducer. Given the extended duration of the experiments spanning several days and the potential for slight alterations in fracture morphology under high-pressure flow conditions, we measured the absolute permeability prior to each set of experiments with different gases and used it for obtaining relative permeability. The measured permeability values were 23 D for hydrogen - Run1 and 8.5 D for hydrogen - Run2 , 7.7 D for methane, and 3.5 D for nitrogen. This protocol ensures that relative permeability calculations for each gas system are normalized against its corresponding fracture geometry which minimizes the effect of fracture evolution. While absolute permeability variations between gas systems indicate some morphological changes occurred between experiments, the static boundary condition assumption remains valid within each individual relative permeability experiment. Furthermore, such changes in absolute permeability have been shown to have minimal impact on the shape of relative permeability curves \cite{Eliebid2024}}.
 \begin{equation}
     Q = {-\frac {K A}{ \mu}} \frac {dP}{L}
\label{permeability}
 \end{equation}
where, $Q$ is volumetric flow rate, $K$ is absolute permeability, $A$ is cross-sectional area of fracture, $\mu$ is viscosity of the fluid injected, dP is differential pressure across the sample and $L$ is the length of sample.

Figure \ref{Aperture} shows the fracture aperture distribution prior to the experiments. We did not observe any visible change in the fracture geometry in the images after the experiments. However, the measured permeability did vary, likely due to subtle pore-scale alterations that lead to fine migration during flow. Since permeability is a macroscopic property, even minor rearrangements at the pore-scale can significantly influence its value. Moreover, the image resolution is insufficient to capture voxel-scale changes in fracture geometry.

\item{The fractional flow of gas is increased from 0.1 to 1 in stepwise manner. For each fractional flow, sample is scanned after steady-state is reached. The steady-state is confirmed by constant differential pressure across the length of rock sample. Fractional flow of gas is defined by equation \ref{fractionaflow}. The capillary number, $Ca_t$ for each fractional flow is defined as equation \ref{Ca_definition} \cite{Spurin2019a} and is given in table \ref{Ca} for each experiment in this study.}

 \begin{equation}
     f_g = {\frac {Q_g}{Q_g + Q_w} }
\label{fractionaflow}
 \end{equation}

where, f$_g$ is gas fractional flow, Q$_g$ is gas volumetric flow rate and Q$_w$ is brine volumetric flow rate. 

 \begin{equation}
     Ca_t = {\frac {Q_t}{\sigma (\frac{f_g}{\mu_{nw}} + \frac{1-f_g}{\mu_w})}}
\label{Ca_definition}
 \end{equation}

where, Ca$_t$ is total capillary number at fractional flow during co-injection, Q$_t$ is total volumetric flow rate (Q$_g$ + Q$_w$), $\mu_{nw}$ is viscosity of non-wetting phase and $\mu_w$  is viscosity of wetting phase. 

\item{Relative permeability: Relative permeability is measured at each fractional flow using equation \ref{relperm} \cite{Jackson2018}.}

 \begin{equation}
     Q_i = {-\frac {K K_{r,i}(S_i) A}{ \mu_i}} \frac {dP}{L}
\label{relperm}
 \end{equation}

where, Q$_i$ is the volumetric flow rate of phase $i$, k is absolute permeability, K$_{r,i}$(S$_i$) is the relative permeability of phase $i$, A is the cross-sectional area of fracture, $\mu_i$ is the viscosity of phase $i$ and dP/L is the pressure gradient.

\begin{table}
 \caption{Capillary number for each fractional flow for different gas-brine systems.}
\centering
 \begin{tabular}{ c  c   c  c  c p{2cm} }
\hline
  f$_g$ &  q$_g$(ml min$^{-1}$)   &  q$_w$(ml min$^{-1}$)  &                      & Ca$_t$ x 10$^{-7}$    &                  \\
 \hline
           &             &               & H$_2$ - brine & CH$_4$ - brine             & N$_2$ - brine  \\
  \hline          
     0.1 &   0.01   & 0.09      & 2.36  & 4.8  & 17.2   \\
     0.3 &   0.03   & 0.07      &  0.84  & 1.76 & 10.28 \\
     0.6 &   0.06   & 0.04      &  0.43 & 0.91   & 6.41   \\
     0.8 &   0.08   & 0.02      &   0.32  & 0.68   & 5.12 \\
     1.0 &   0.1     & 0           &  0.26  & 0.55  & 4.27  \\
  \hline
\label{Ca}
 \end{tabular}
\end{table}

\end{enumerate}

\subsection{Imaging and image analysis} \label{ssec:Imaging}

X-ray imaging was performed with the `High Energy micro-CT Optimized for Research' scanner (HECTOR) \cite{Masschaele2013} at the center for X-ray tomography at Ghent University (UGCT). The X-ray energy and power were 160 kV and 15 W, respectively. We acquired 1301 projections with 600 ms  exposure time for each projection. For each fractional flow at steady state entire sample length is scanned with imaging field of view (FOV) 2000 x 2400 x 1400 voxels. The voxel size was \SI{30.37}{\micro\metre} for each image. The Octopus software (Tescan-XRE, Belgium) was used to reconstruct the 3D image from acquired projections.  

The reconstructed images were processed using Avizo (ThermoFisher scientific) software. All the images from wet/fractional flow scans were registered to the dry scan using normalized mutual information and denoised using non-local means filter with similarity value of 0.4. The fracture from the dry scan was segmented using global thresholding, which was then applied as a mask for segmenting the wet or fractional flow scans. Two fluid phases, brine and gas, were segmented from the masked wet images using histogram-based thresholding. The total porosity of the fracture was determined from the volume fraction of the segmented dry scan, while the saturation of each fluid phase was calculated along the length of the rock sample from the segmented wet scans.

\section{Results} \label{sec:results}

\subsection{Relative permeability} \label{ssec:relperm}

\begin{figure}[H] 
\centering
\begin{subfigure}[b]{0.49\textwidth}
 \centering
 \includegraphics[width=\textwidth]{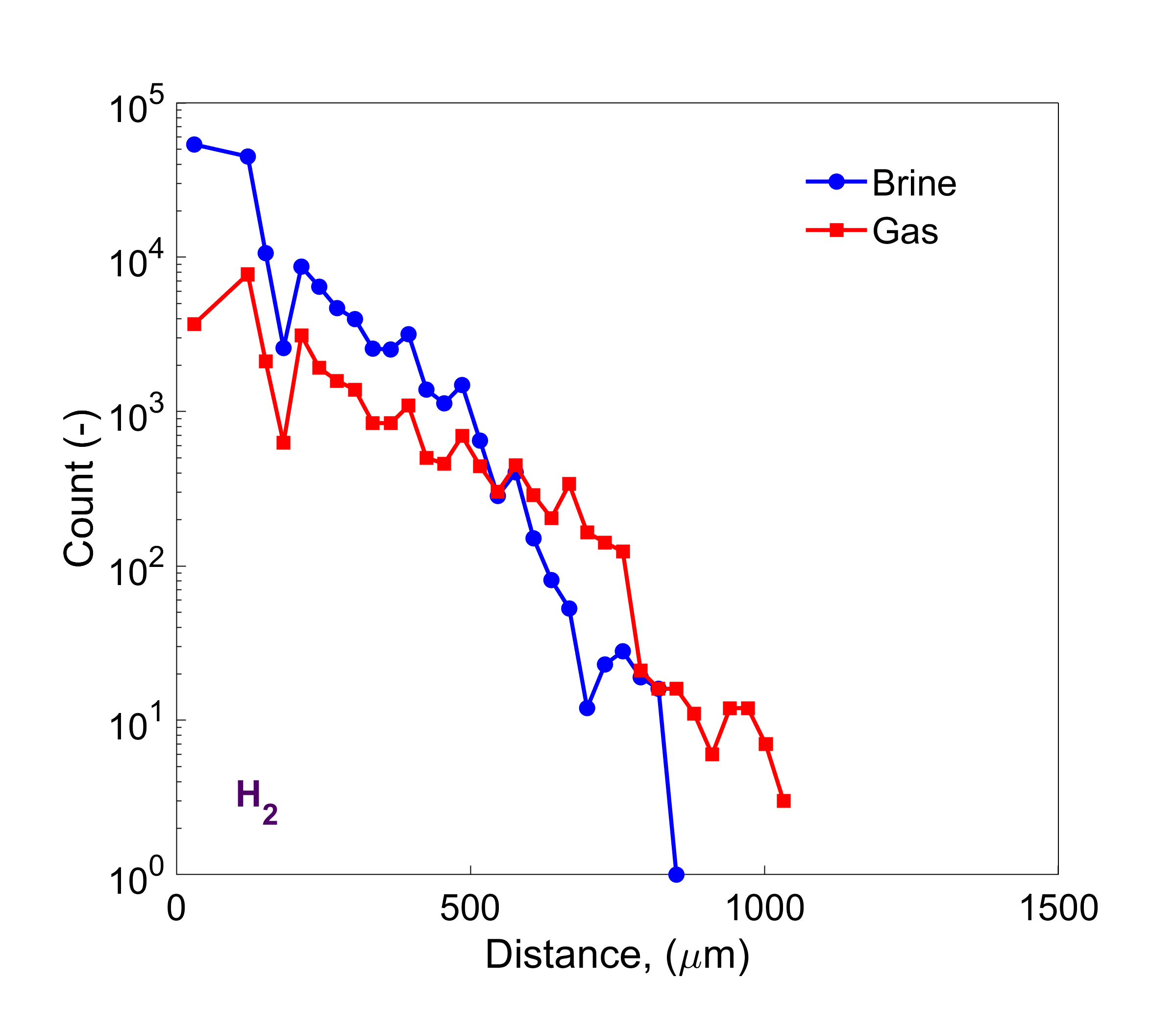} 
   \caption{}
\label{Occupancy_H2repeat}
\end{subfigure}
\begin{subfigure}{0.49\textwidth}
 \centering
 \includegraphics[width=\textwidth]{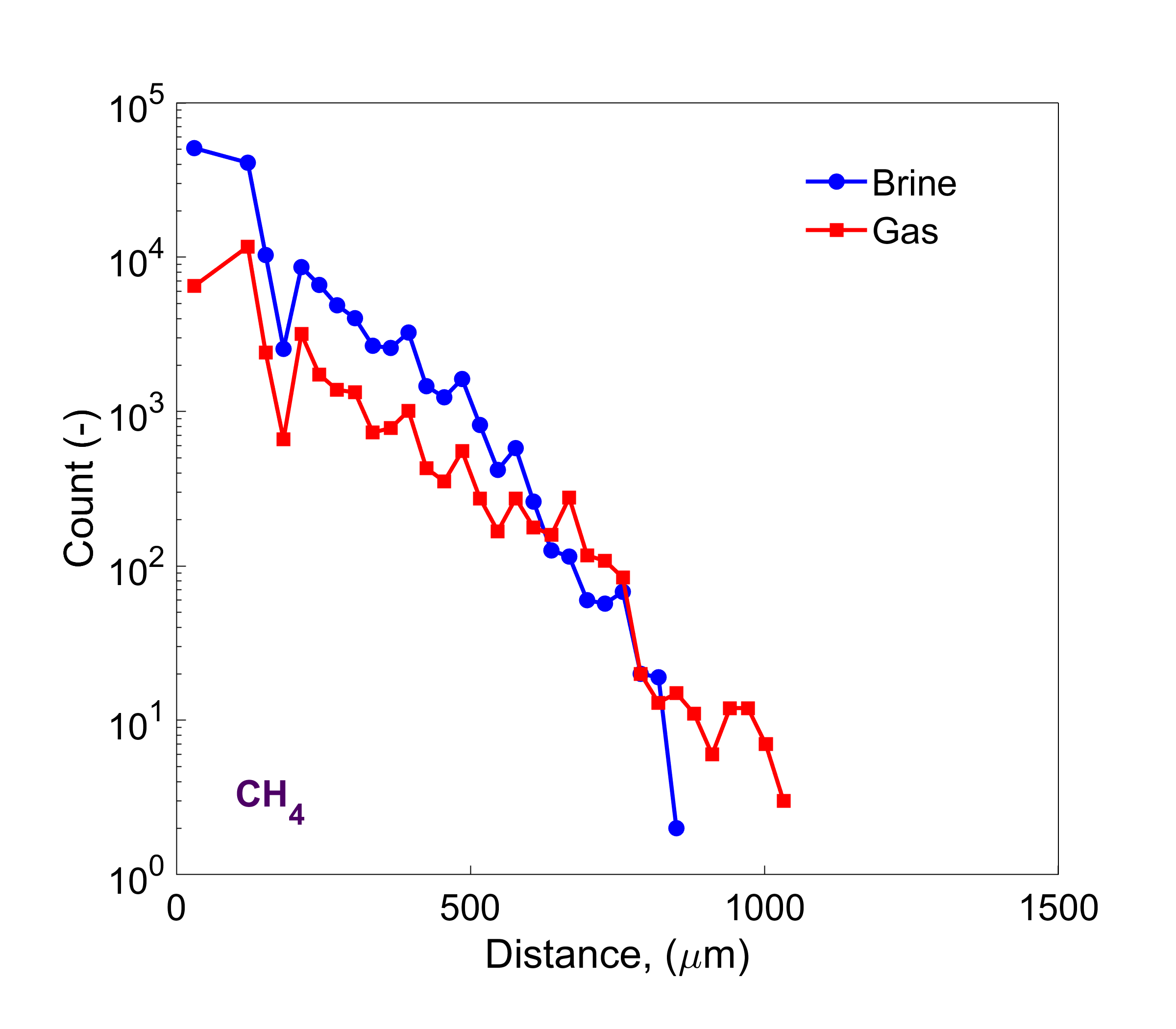} 
  \caption{}
\label{Occupancy_CH4}
\end{subfigure}
\begin{subfigure}{0.49\textwidth}
 \centering
 \includegraphics[width=\textwidth]{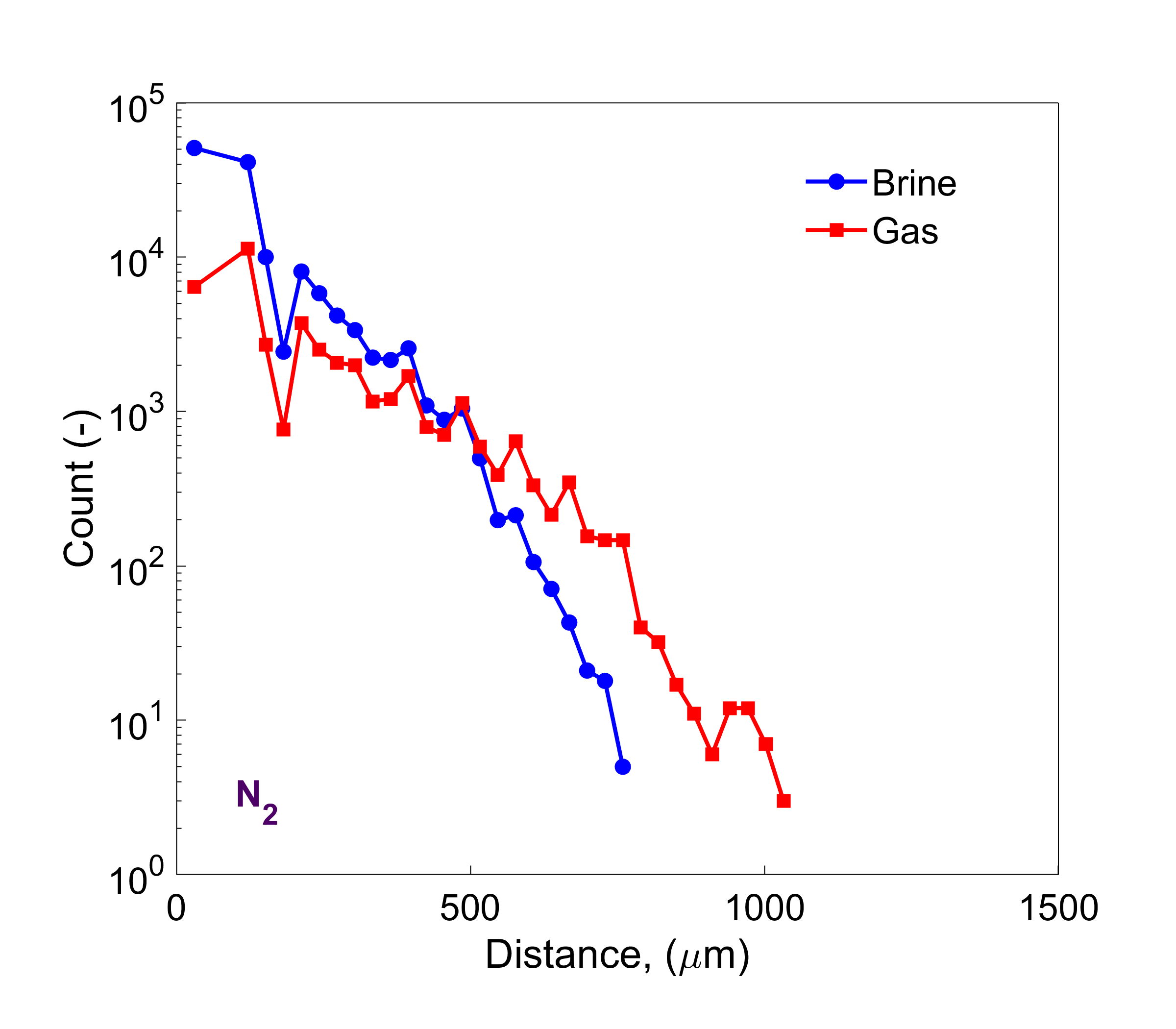} 
  \caption{}
\label{Occupancy_N2}
\end{subfigure}
\caption{Fluid occupancy at $f_g = 0.6$ for different gases: (a) H$_2$ (b) CH$_4$ (c) N$_2$. The distance is derived from the distance map of the segmented dry fracture image, calculated as twice the distance between the center and the nearest wall of the fracture.}
\label{Fluid_Occupancy_F3}
\end{figure}

\begin{figure}[h]
 \centering
 \noindent\includegraphics[width=\textwidth]{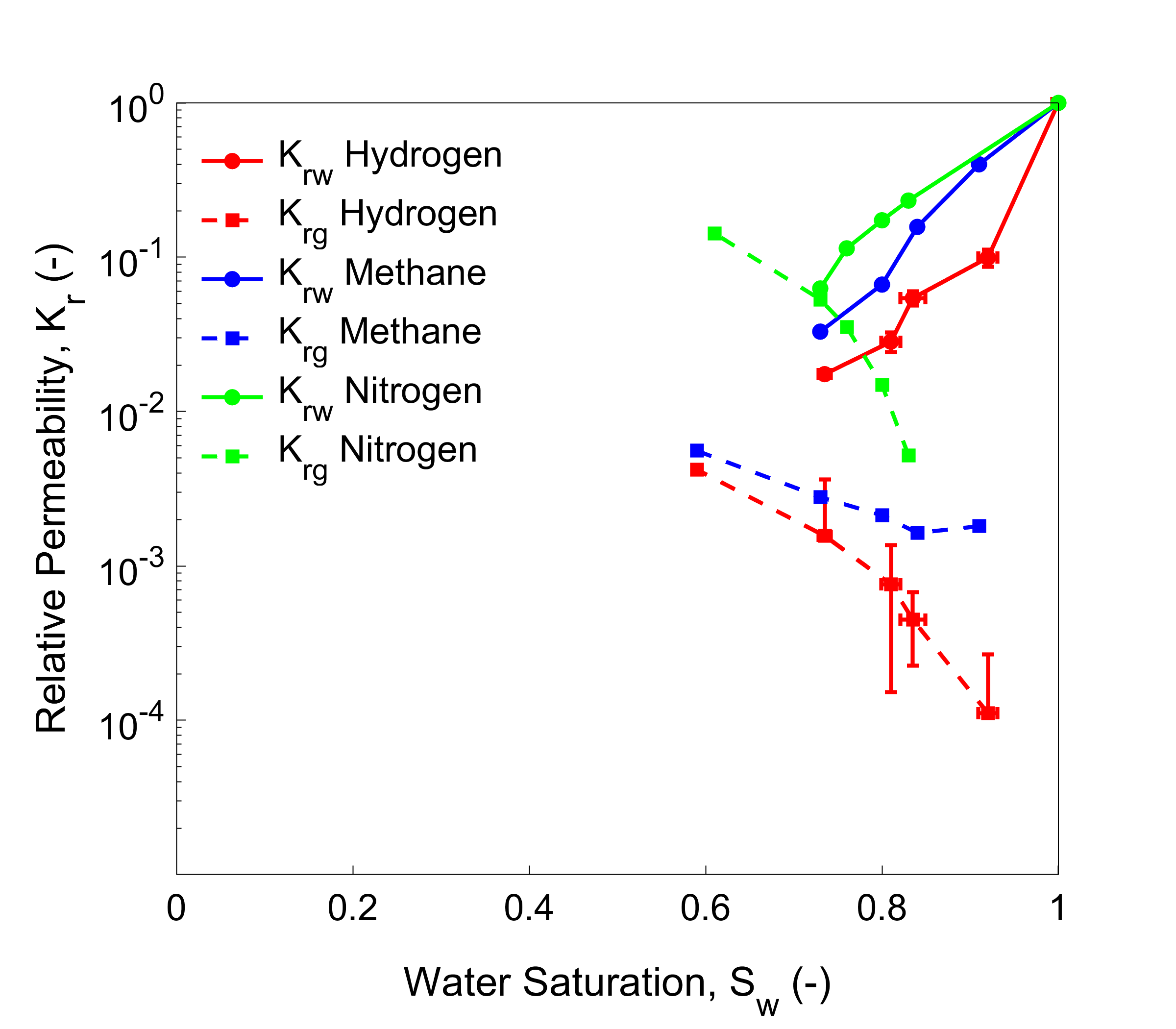}
  \caption{Drainage steady-state relative permeability comparison between H$_2$-brine, CH$_4$-brine and N$_2$-brine system at 10 MPa and 22$^{\circ}$C.}
 \label{RelPerm}
\end{figure}

The relative permeability curves were derived directly from stabilized differential pressure and saturation data, as shown in Figure \ref{RelPerm}. During drainage, as the fractional flow of gas ($f_g$) increases and the capillary number decreases, the system moves to lower water saturations, resulting in better-connected gas-phase displacement. This leads to an increase in gas relative permeability ($K_{rg}$) and a decrease in brine relative permeability ($K_{rw}$). Previous studies report that hydrogen-water systems are water-wet with respect to calcite \cite{Aghaei2023}. In such water-wet systems, gas preferentially invades regions with lower capillary entry pressures during drainage, corresponding to fracture spaces with larger apertures \cite{Karpyn2007, Phillips2023, Chen2017}. Figure \ref{Fluid_Occupancy_F3} illustrates gas-brine occupancy at $f_g = 0.6$ for hydrogen, methane, and nitrogen, where gas is observed occupying largest aperture spaces, while brine prefers to remain in narrower regions. This confirms that this system is water-wet. It it  consistent with previous observations in porous media, where gas invades larger pores and brine remains in narrower throats \cite{Wang2023a}. Similar trends were observed at other fractional flows, as shown in the Supplementary Information S1.

The relative permeability-saturation curves for the hydrogen-brine and methane-brine systems show closely aligned trends and similar endpoint saturations. Both gases exhibit low gas relative permeability ($K_{rg}$), indicating substantial phase interference caused by the heterogeneous fracture geometry (roughness) and the low viscosity ratios of the gas-brine systems (0.01 for hydrogen-brine and 0.016 for methane-brine). This behaviour suggests the occurrence of intermittent flow and snap-off events, further discussed in Section \ref{ssec:pressure_analysis}. The wetting-phase permeability ($K_{rw}$) consistently remains much higher than the gas permeability, indicating that brine stays well-connected across all fractional flows. This strong water-wet condition ensures brine continuity throughout the fracture network, even as gas saturation increases. At higher water saturation (lower fractional flows, $f_g$), the gas relative permeability ($K_{rg}$) is slightly higher for the methane-brine system compared to the hydrogen-brine system. This is attributed to the higher capillary number (Ca$_t$) at high wetting phase saturation, where viscous forces contribute more to the flow dynamics. The slightly higher viscosity ratio of the methane-brine system (0.016 versus 0.01 for hydrogen-brine) facilitates better connectivity in the methane front, resulting in higher gas permeability \cite{Spurin2019a}. However, as the water saturation decreases (e.g., $f_g = 0.8$ and $f_g = 1$), capillary forces dominate, leading the $K_{rg}$ values for hydrogen and methane to converge at similar endpoint saturations $S_w = 0.6$ for hydrogen and $S_w = 0.59$ for methane. These slight differences in relative permeability and saturation at endpoints may fall within the experimental margin of error. Notably, the saturation change from $f_g = 0.1$ to $f_g = 1$ is identical for both gases, decreasing from 0.91 to 0.6. It should be noted that the entire hydrogen-brine relative permeability experiment was repeated to assess the uncertainty in the measurements and will be discussed in \ref{ssec:sensitivity_analysis}.

The nitrogen-brine relative permeability curves show a significant difference from those of hydrogen-brine and methane-brine, specifically with nitrogen maintaining higher gas ($K_{rg}$) and water permeability ($K_{rw}$) across all fractional flows. At the endpoint where water saturation ($S_w$) converges to 0.6 for all gases, the gas relative permeability ($K_{rg}$) reaches 0.14 for nitrogen, while it is notably lower for hydrogen and methane, at 0.007 and 0.0055, respectively. This difference arises because nitrogen's higher viscosity, an order of magnitude greater than that of hydrogen and methane, increases the capillary number under the same volumetric flow rate conditions. At 10 MPa, nitrogen exists in a supercritical state, which affects its fluid behavior in the fracture system. At the initial fractional flow of $f_g = 0.1$, the wetting-phase saturation for nitrogen is notably lower than that of hydrogen or methane.


\subsection{Saturation and fluid distribution} \label{ssec:saturation}

\begin{figure}[H] 
\centering
\begin{subfigure}[b]{0.49\textwidth}
 \centering
 \includegraphics[width=\textwidth]{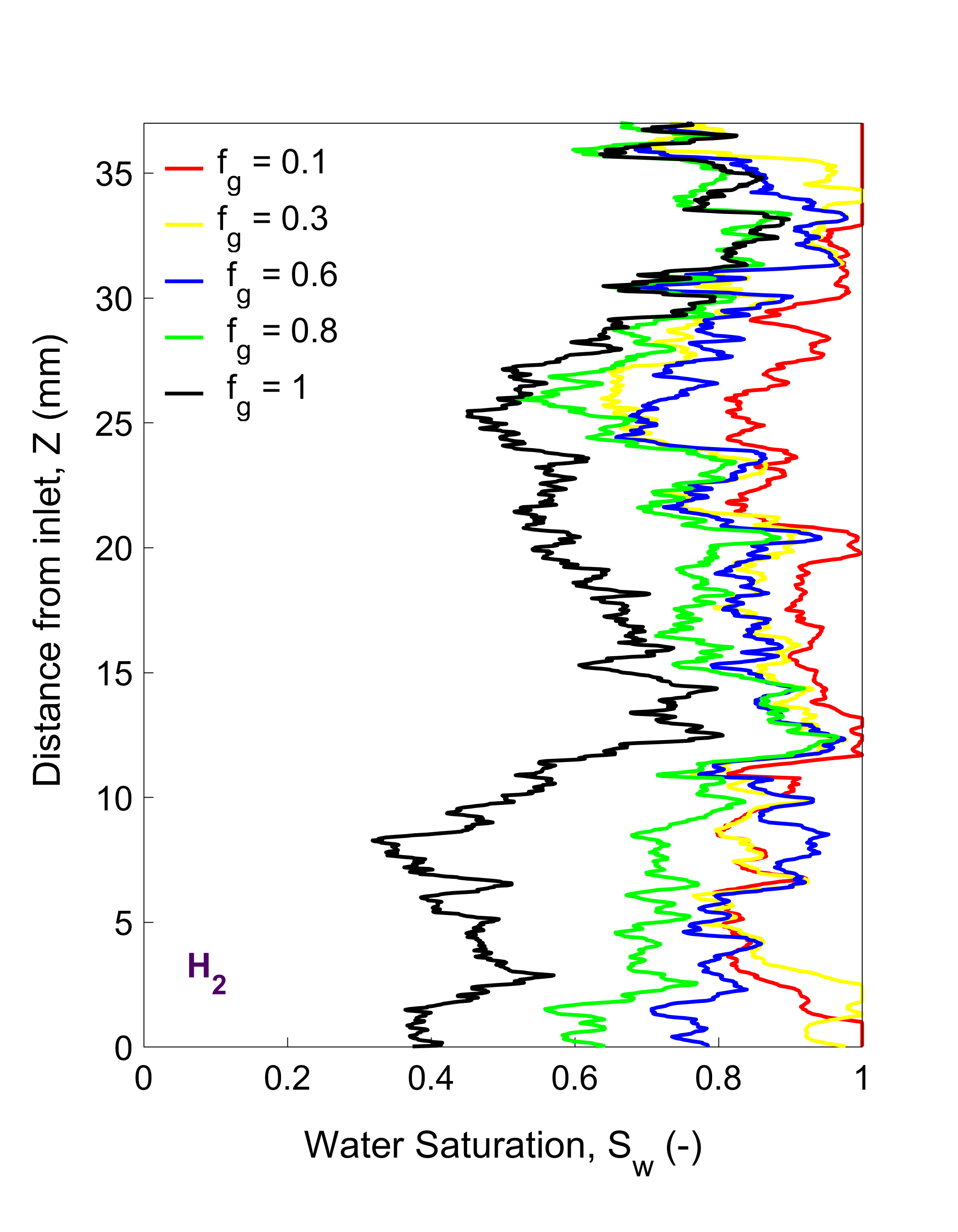} 
   \caption{}
\label{Sat_H2}
\end{subfigure}
\begin{subfigure}[b]{0.49\textwidth}
 \centering
 \includegraphics[width=\textwidth]{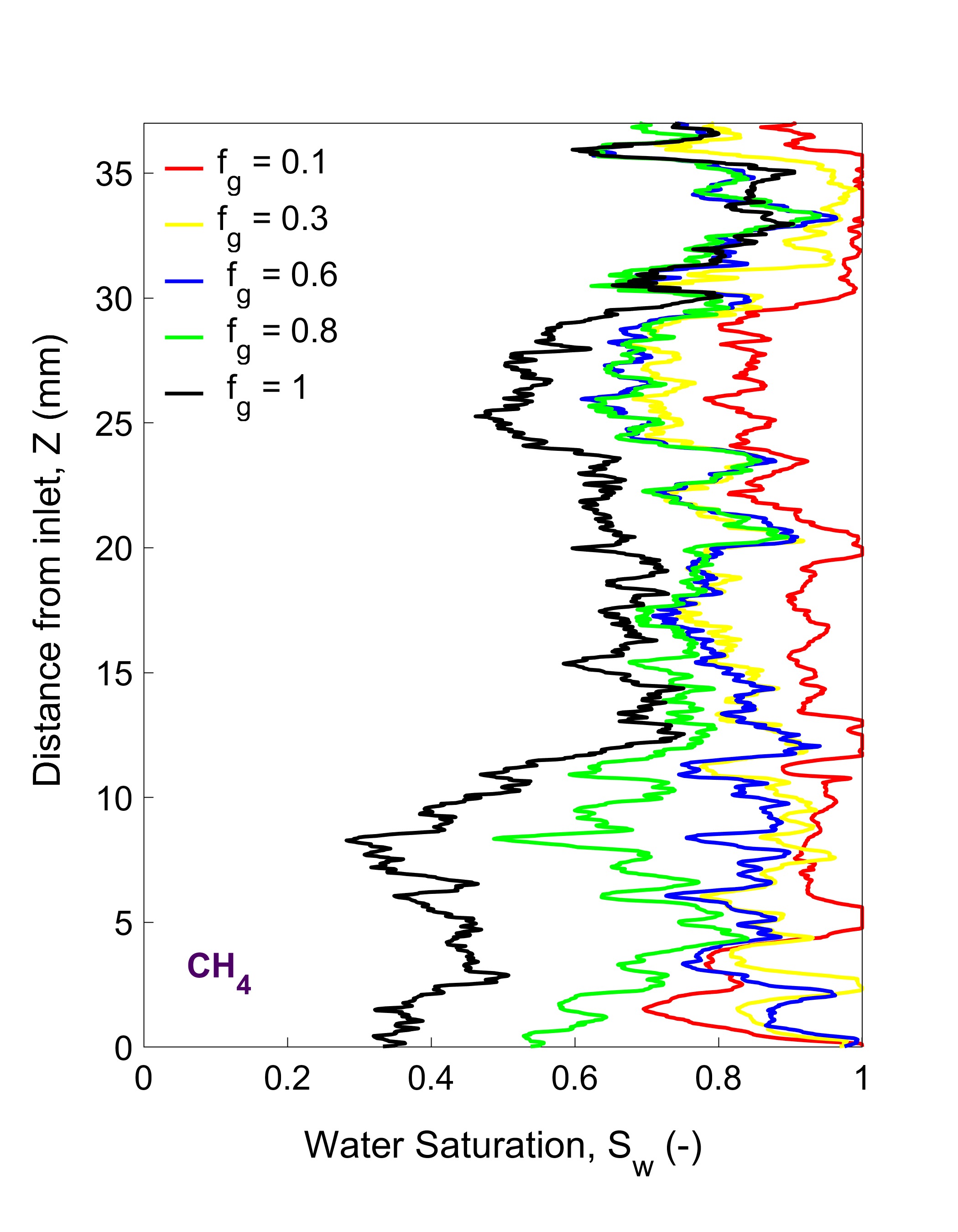} 
   \caption{}
\label{Sat_CH4}
\end{subfigure}
\begin{subfigure}{0.49\textwidth}
 \centering
 \includegraphics[width=\textwidth]{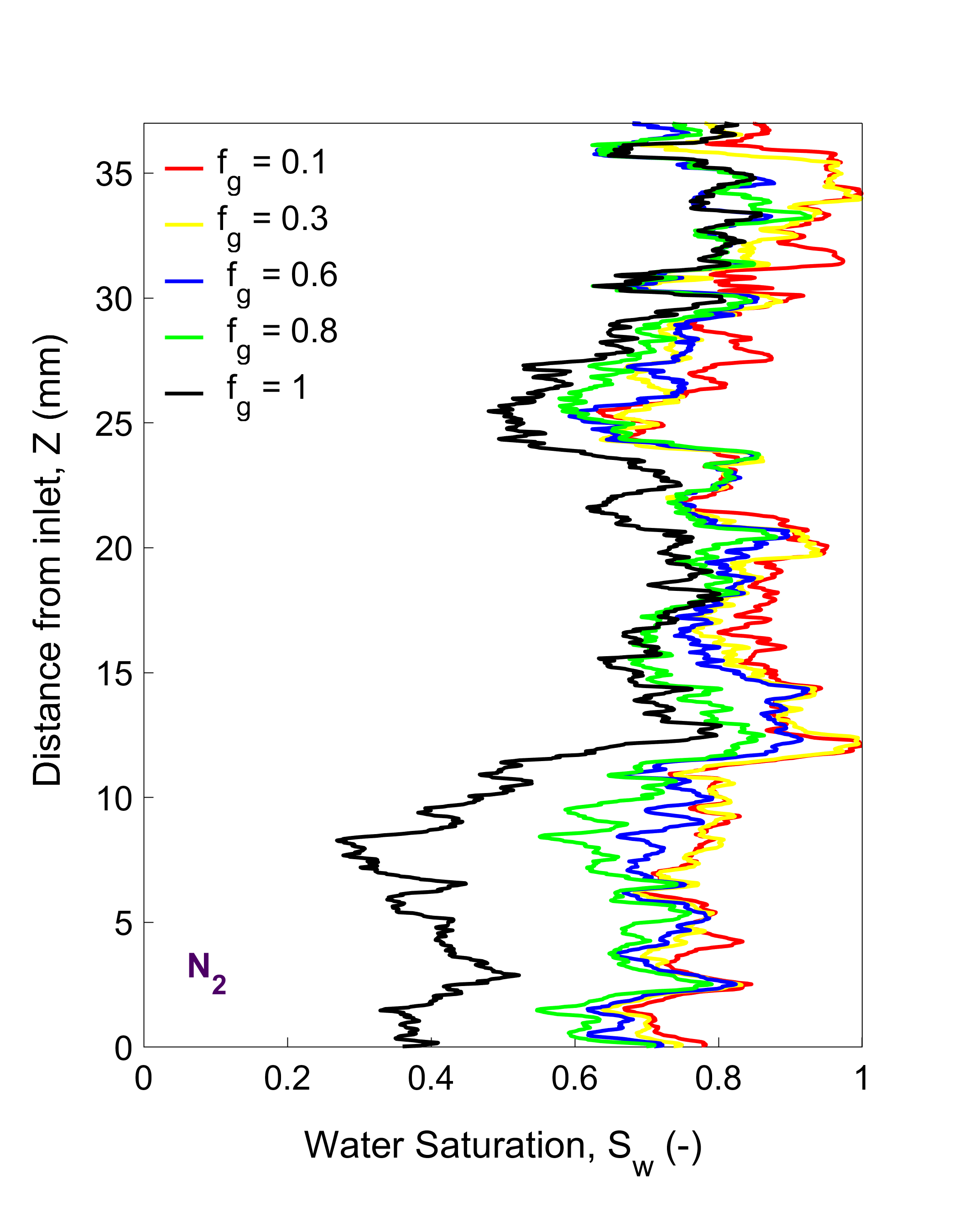} 
  \caption{}
\label{Sat_N2}
\end{subfigure}
\caption{1-D slice-averaged saturation profile of brine along the length of the core for different gases. (a) H$_2$ (b) CH$_4$ (c) N$_2$.}
\label{Saturation_H2_CH4_N2}
\end{figure}

To further explore the cause of the similarities and differences between the different relative permeability measurements, we investigated the average saturation and fluid distribution in the fracture space. The slice-averaged 1-D saturation profiles for hydrogen, methane, and nitrogen experiments at different fractional flows are shown in Figures \ref{Sat_H2}, \ref{Sat_CH4}, and \ref{Sat_N2}, respectively. As the fractional flow of gas ($f_g$) increases, the slice-averaged brine saturation (S$_w$) across the sample decreases for all gases, indicating that more gas is occupying the fracture space. For hydrogen and methane, saturation varies between 0.3 and 1, while for nitrogen it ranges from 0.27 to 0.99, suggesting a slightly lower average saturation for nitrogen-brine system. It results from the fracture geometry being highly heterogeneous with variable aperture distribution and roughness as shown in Figure \ref{Aperture}.

\begin{figure}[H] 
\centering
\begin{subfigure}[b]{0.49\textwidth}
 \centering
 \includegraphics[width=\textwidth]{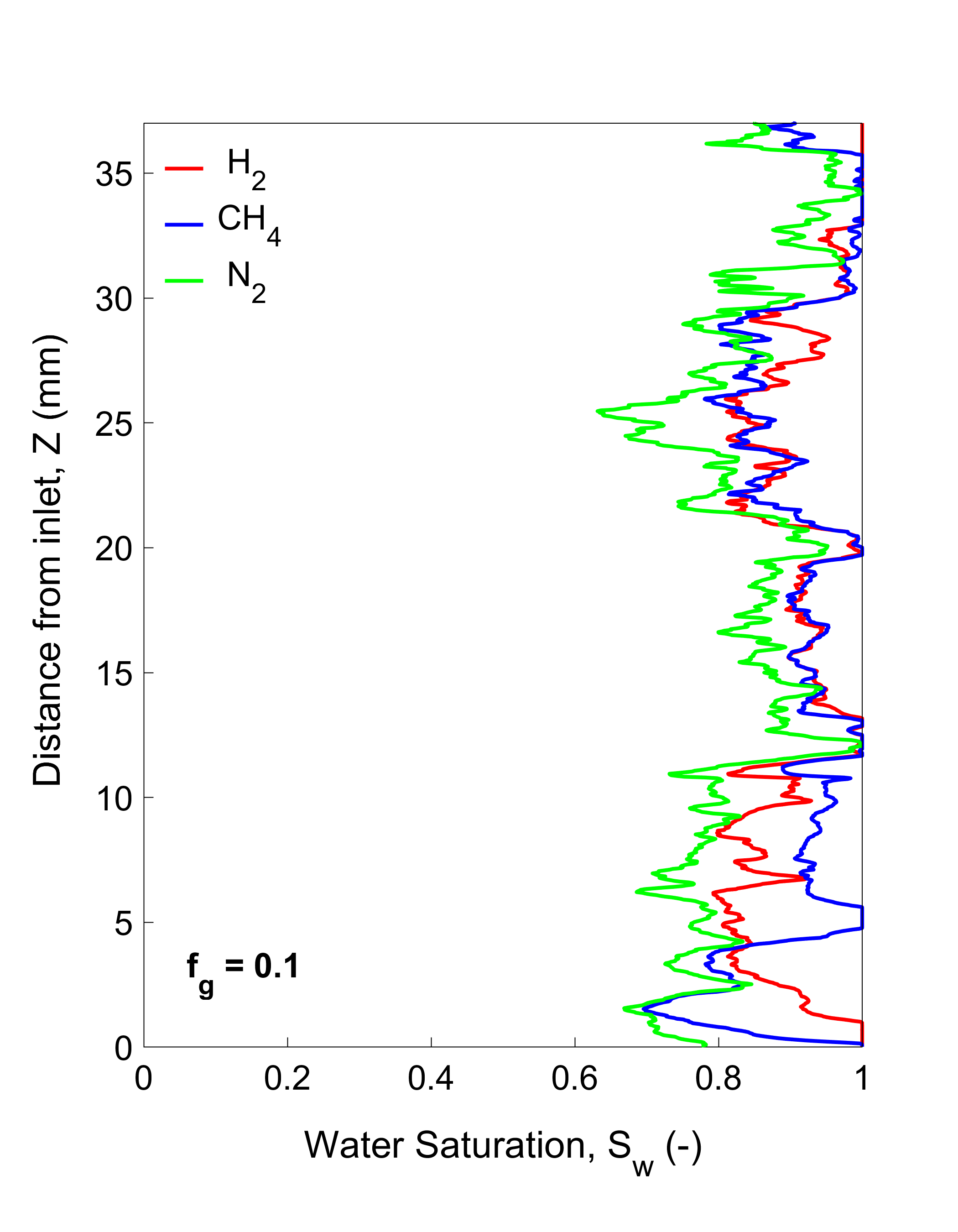} 
   \caption{}
\label{Sat_fg01}
\end{subfigure}
\begin{subfigure}{0.49\textwidth}
 \centering
 \includegraphics[width=\textwidth]{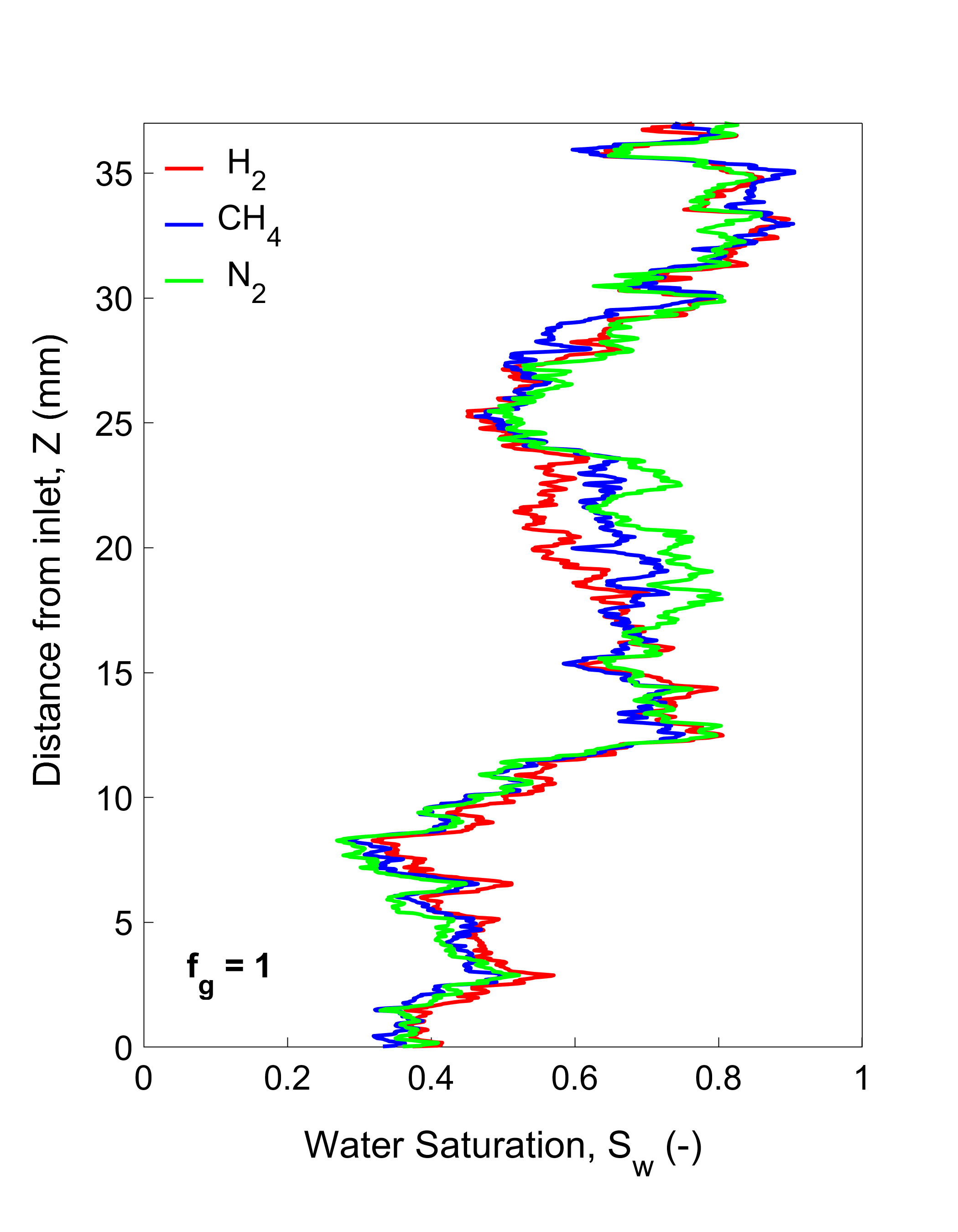} 
  \caption{}
\label{Sat_fg1}
\end{subfigure}
\caption{1-D slice-averaged saturation profile of brine along the length of the core for different gases.(a) $f_g$ = 0.1 (b) $f_g$ = 1.}
\label{Saturation_different gases}
\end{figure}

\begin{figure}[H]
 \centering
 \noindent\includegraphics[width=\textwidth]{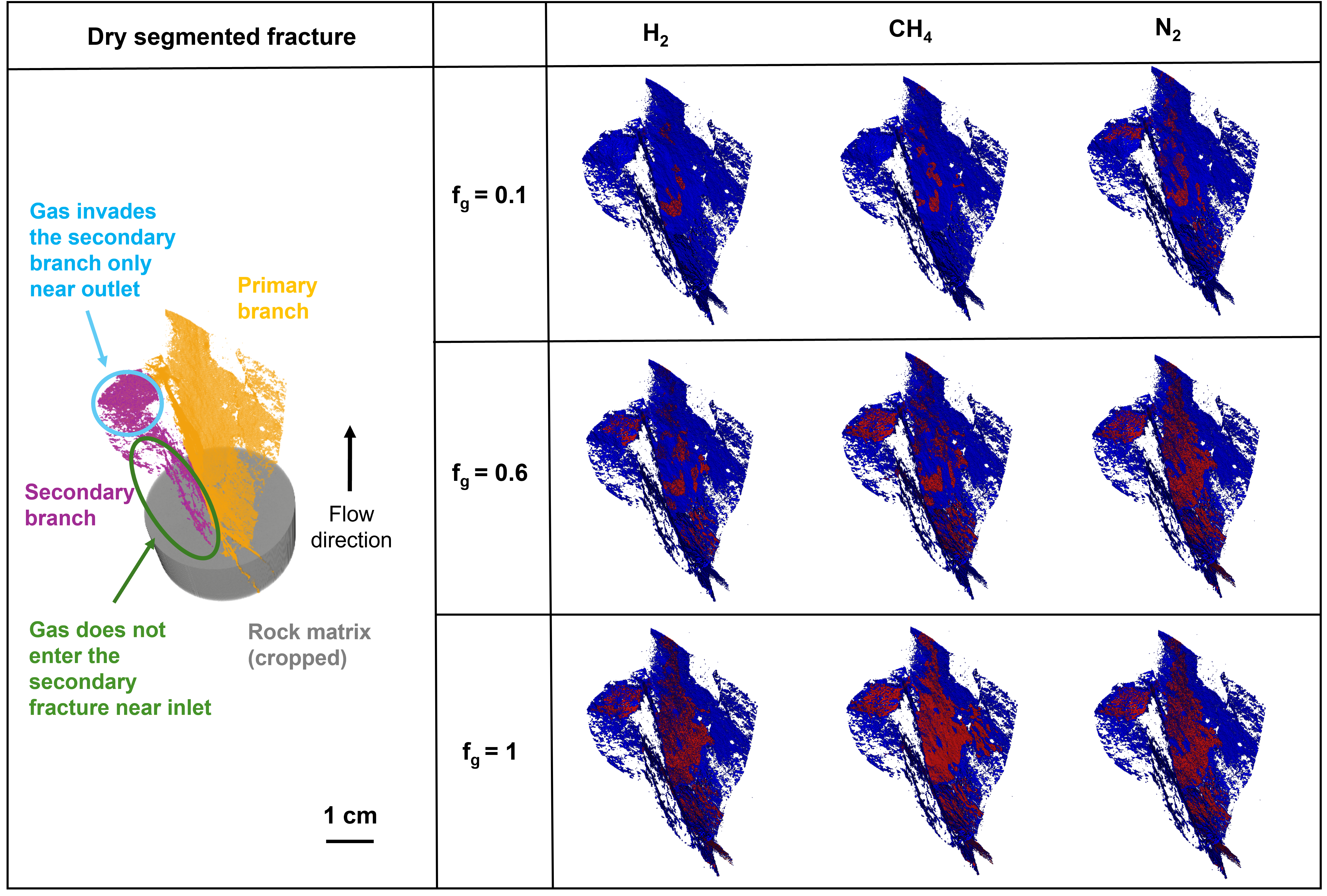}
 \caption{(left) 3D volume rendering of the fracture. The orange color denotes primary fracture and purple color denotes the secondary fracture. (right)  Fluid distribution in the fracture at three different fractional flows. The horizontal panels shows variation for different gases at the same fractional flow and vertical panels show variation in fractional flows for the same gas type. The red color denotes gas and blue color denotes brine phase.}
\label{FluidDistribution_F1F3F5}
 \end{figure}

At lowest fractional flows $f_g$ = 0.1, the saturation profiles illustrated in Figure \ref{Sat_fg01} shows noticeable differences, with brine saturation for hydrogen being 5$\%$ higher than for methane and 10$\%$ higher than for nitrogen. This variation primarily results from hydrogen's lower viscosity, which leads to less efficient invasion of the fractures, ultimately resulting in lower gas saturation and higher wetting-phase saturation. Although the fluid properties of hydrogen and methane at 10 MPa and 22$^{\circ}$C are similar, nitrogen, in its supercritical state, demonstrates distinctly different flow behavior. The higher viscosity of nitrogen facilitates more efficient invasion of the fracture network, resulting in higher gas saturation. Nitrogen is observed to drain more efficiently compared to hydrogen and methane in porous media, as also reported by \citet{Higgs2023}. Despite the interfacial tension (IFT) of nitrogen (0.07 N m$^{-1}$) and hydrogen (0.069 N m$^{-1}$) being comparable at 10 MPa, the differences in viscosity significantly influence wetting-phase saturation at onset of drainage. This underscores the finding that at low fractional flows, the viscosity ratio plays a more pronounced role in gas displacement behavior. The fluid distribution in the fracture at stable fractional flow, $f_g$ = 0.6  as shown in Figure \ref{FluidDistribution_F1F3F5} (middle horizontal panel), indicates that nitrogen has the highest gas fraction (S$_w$ = 0.76) in the fracture, followed by methane (S$_w$ = 0.79), and then hydrogen (S$_w$ = 0.83 ).  

At higher gas fractional flows ($f_g$ = 1) the saturation converges for all gases as shown in Figure \ref{Sat_fg1}. At these higher gas fractional flows, the capillary number (Ca$_t$)  decreases resulting in more connected gas which diminishes the impact of the viscosity ratio. Consequently,  similar saturations are observed at $f_g$ = 1, representing endpoint saturation. Although the nitrogen case has a significantly higher Ca$_t$ (1.2 x 10$^{-6}$) than hydrogen and methane (7.0 x 10$^{-6}$ and 1.5 x 10$^{-6}$ respectively), the end point saturations remain consistent for all gases at S$_w$ = 0.6. This suggests that the effect of capillary number on overall saturation is relatively minor under the tested conditions.

\begin{figure}[H]
\centering
\begin{subfigure}[b]{0.64\textwidth}
    \centering
    \includegraphics[width=\textwidth]{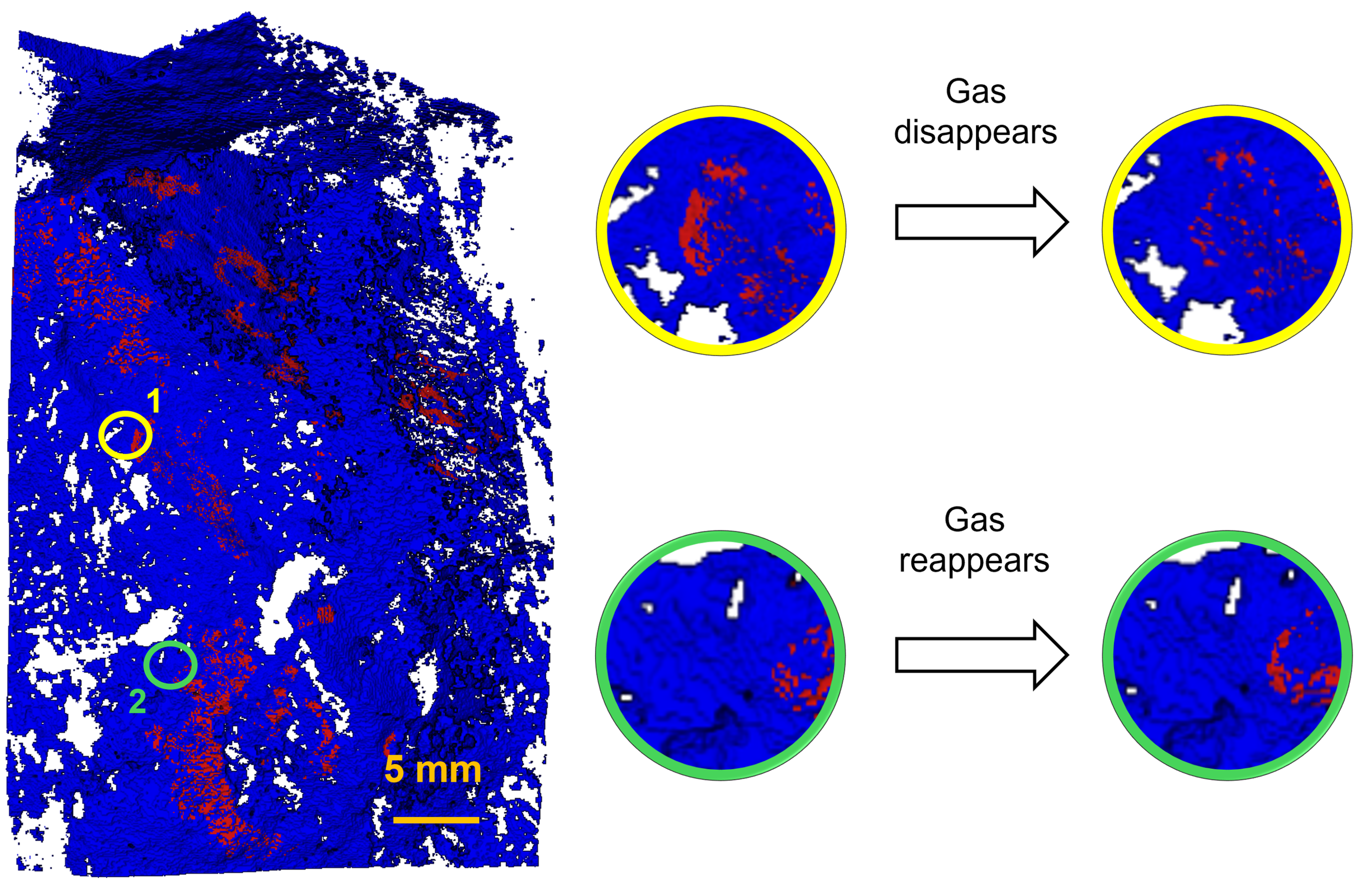}
    \caption{}
    \label{FluidDistribution_intermittency}
\end{subfigure}
\begin{subfigure}[b]{0.35\textwidth}
    \centering
    \includegraphics[width=\textwidth]{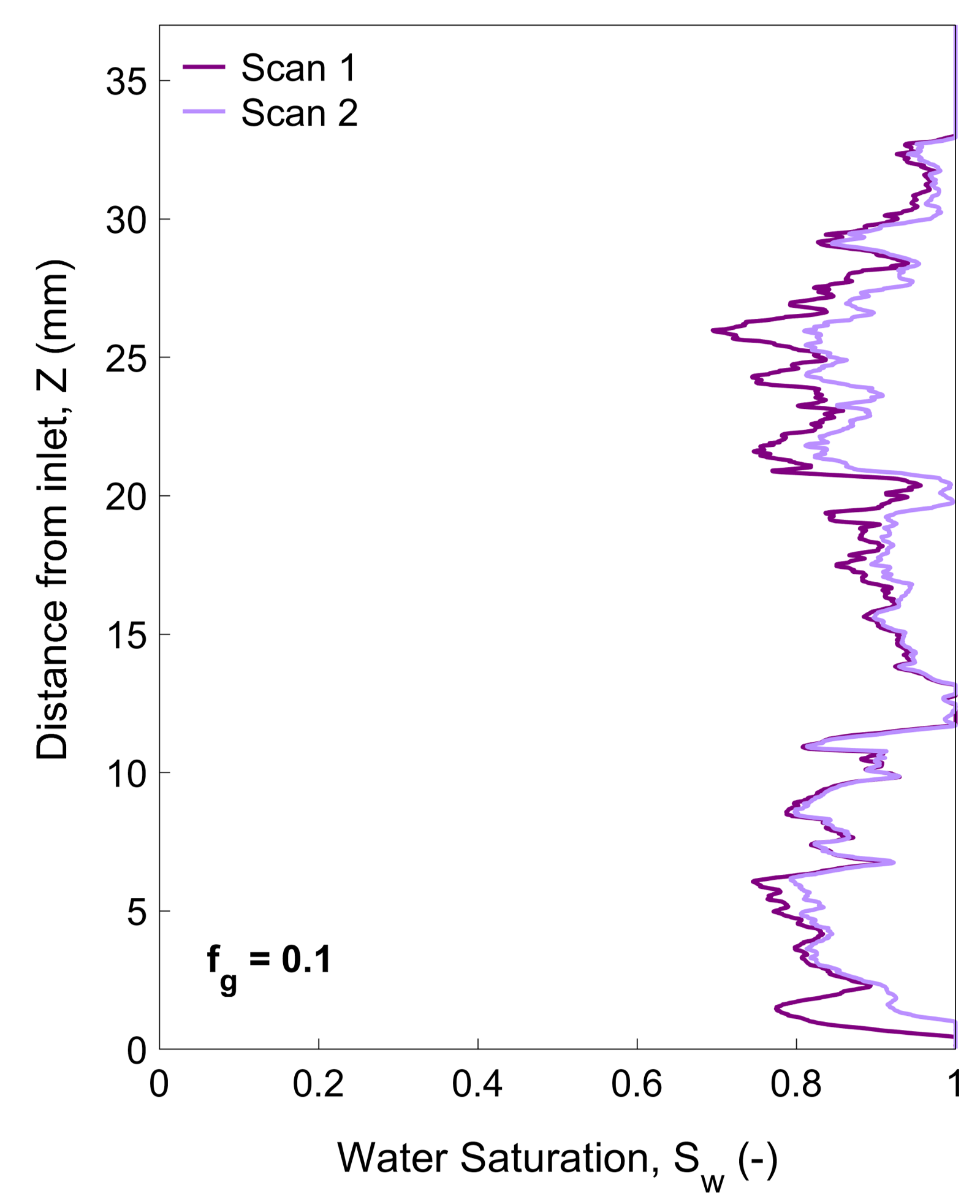}
    \caption{}
    \label{Sat_H2_intermittency_fg01}
\end{subfigure}
\caption{H$_2$ - brine flow at $f_g = 0.1$ for two scans taken 30 minutes apart. (a) 3D fluid distribution and intermittency. The red color denotes hydrogen, and blue denotes brine. (b) 1-D slice-averaged saturation profile.}
\label{30minScans}
\end{figure}

The fluid distribution for the hydrogen-brine system in the fracture exhibits signs of intermittency at higher water saturations (lower $f_g$), as illustrated in Figure \ref{FluidDistribution_intermittency} for $f_g = 0.1$. Subtle differences in gas distribution are observed between two scans taken 30 minutes apart after reaching steady-state pressure. Several locations are identified by coloured circles: Circle 1 (yellow) marks a region where gas disappears in the subsequent scan, while Circle 2 (green) indicate location where gas reappears. These fluctuations in the gas phase correspond to capillary events and are also reflected in the measured differential pressure. In contrast, the nitrogen-brine system exhibits fewer fluctuations in saturation, suggesting a more stable flow pattern. A more detailed analysis of pressure fluctuations, intermittency, and their impact on fluid distribution will be discussed in Section \ref{ssec:pressure_analysis}. Along the length of the fracture, slice-averaged saturation reaches S$_w$ = 1 for many slices, indicating a disconnected gas phase. Figure \ref{Sat_H2_intermittency_fg01} presents the 1-D slice-averaged saturation profiles for these two subsequent scans revealing significant fluctuations in saturation within this period, emphasizing dynamic capillary processes occurring in the fracture. These processes are not fully captured in terms of intermediate gray values due to the scanner's limited temporal resolution. As the gas fractional flow increases to $f_g$ = 1, the gas phase becomes more connected throughout the fracture, which reduces intermittency and enhances overall gas saturation as demonstrated by the 3D volume rendering of fluid distribution in Figure \ref{FluidDistribution_F1F3F5} (bottom horizontal panel). Intermittency significantly impacts relative permeability, resulting in notably low gas permeabilities, particularly for hydrogen and methane as discussed in Section \ref{ssec:relperm}. 

We note here that the observed intermittency may also be influenced by the dissolution of hydrogen and methane into the brine during the experiment, as the injected gas and brine were not pre-equilibrated, a phenomenon previously described by \cite{Boon2022}. While bulk solubilities (hydrogen: 1.6 mg L$^{-1}$, methane: 23 mg L$^{-1}$, and nitrogen 19 mg L$^{-1}$ \cite{Kaye}) suggest hydrogen dissolution is minimal, but pore-scale dissolution kinetics governed by faster diffusion can still cause significant dissolution. However, the reappearance of gas in certain fracture locations indicates that intermittency is a genuine effect and not solely attributed to dissolution.

In addition to flow intermittency, geometric complexities may also contribute to the low gas permeabilities observed in our measurements. The dry scan (Figure \ref{FluidDistribution_F1F3F5} (left)) reveals secondary fractures branching from the central region of the main fracture, forming an intricate fracture network. As we discuss next, these geometric features influence fluid distribution and gas displacement efficiency, thereby affecting overall flow dynamics. The primary branch is highlighted in orange, while the secondary branch is shown in purple. The secondary branch is further subdivided into two regions based on aperture distribution, indicated by light blue and green circles.  No gas invasion is observed near the inlet of the secondary branch (green region) for any gas or fractional flow, indicating a high capillary entry pressure due to the narrow aperture. At higher fractional flows ($f_g$ = 0.6 and $f_g$ = 1) as shown in Figure \ref{FluidDistribution_F1F3F5} (middle and bottom horizontal panel, respectively), gas is present for all gas types in the secondary fracture near the outlet (ligh blue region), where the aperture is wider and better connected. However, at lower fractional flows (f$_g$ = 0.1) (Figure \ref{FluidDistribution_F1F3F5}, top horizontal panel) particularly for low-viscosity gases like hydrogen and methane, gas does not invade this location, suggesting limited gas connectivity in fracture networks under these conditions.

\subsection{Pressure analysis} \label{ssec:pressure_analysis}

\begin{figure}[H]
 \centering
 \noindent\includegraphics[width=\textwidth]{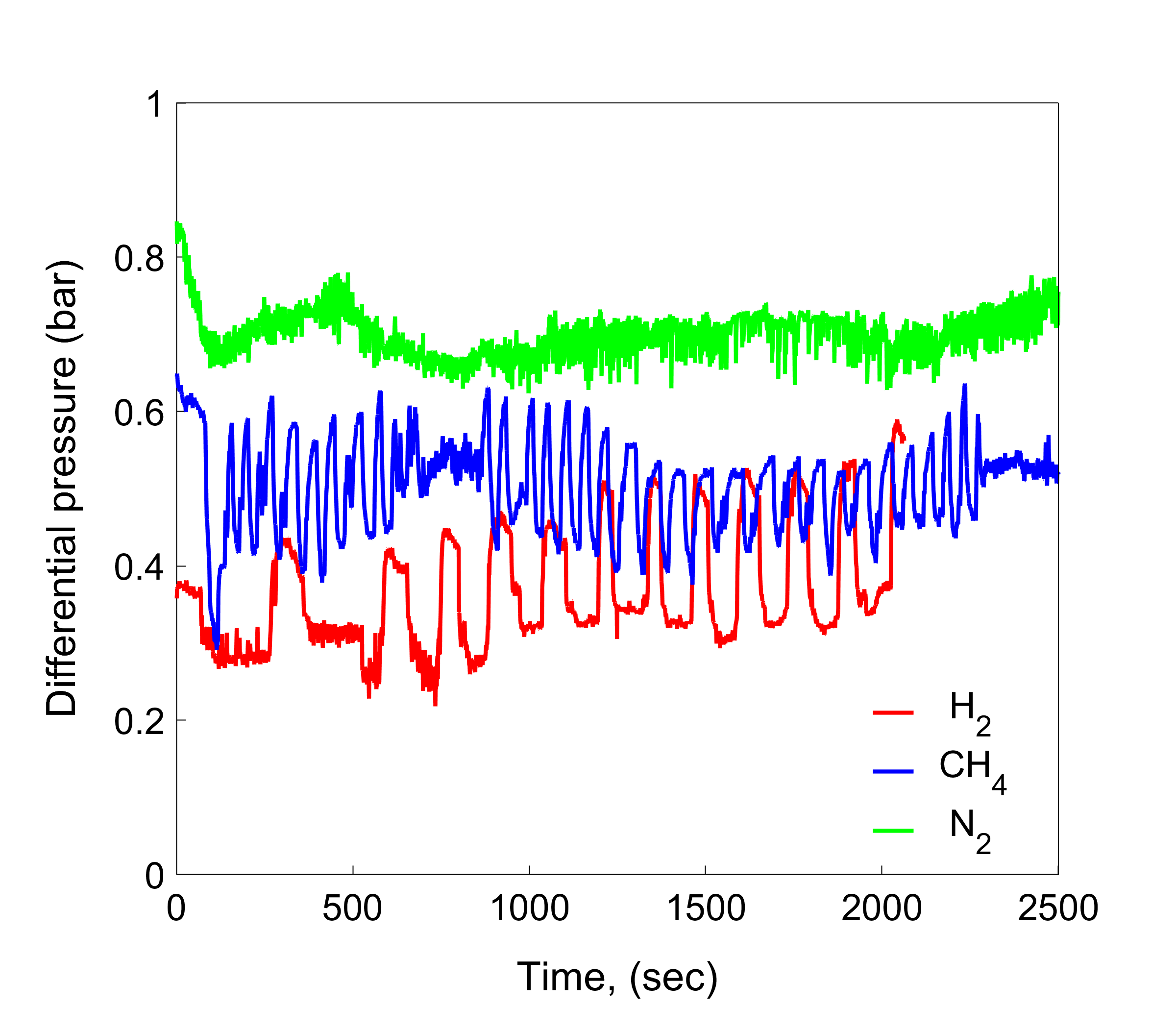}
  \caption{Differential pressure data during steady-state injection of gas and brine for $f_g$ = 0.6.}
\label{Pressuredrop}
 \end{figure}

The lower gas relative permeabilities for similar global water saturation suggest higher phase interference in the fracture. Saturation and fluid distribution in the fracture, based on scans taken 30 minutes apart, indicate the presence of intermittency, where Haines jumps and snap-off events occur. Since the scan duration is 16 minutes and these events appear to happen much more rapidly, we observed pressure data recorded at a much higher time resolution of 1 second. The fluctuations in differential pressure measurements from the transducer across the rock samples provide insights into the timescale and the fluctuations that correspond to snap-off and Haines jumps occurring during two-phase displacements \cite{Spurin2022, Wang2023a}. Differential pressure measurements at $f_g = 0.6$ over a 30-minute period for H$_2$, CH$_4$, and N$_2$ are presented in Figure \ref{Pressuredrop}. These data reveal fluctuations over varied timescales for each gas, underscoring distinct dynamics associated with each gas phase. There are notable differences in the amplitude and frequency of these oscillations, particularly between H$_2$ and CH$_4$ on one hand and N$_2$ on the other. The pressure amplitude of fluctuations is greater for hydrogen and methane compared to nitrogen, with the fluctuations occurring over timescales of approximately 150 seconds for hydrogen, 30 seconds for methane, and one second for nitrogen. The pressure fluctuations within the pores result in intermittent pore filling, occurring over a broad range of timescales. At different fractional flows, the timescales of these fluctuations vary, reflecting changes in the connectivity of the gas phase. The additional pressure data are provided in Supplementary Information S2.

High-speed imaging, such as synchrotron-based X-ray tomography, would offer valuable insights into the pore-scale dynamics of these processes. For example, \citet{Phillips2023} demonstrated that flow in rough fractures exhibits greater phase intermittency than in porous media, even during slow drainage. Additionally, \citet{Spurin2022} validated the correlation between pressure fluctuations and fluid distribution in porous media, highlighting the potential of pressure data as a tool for understanding intermittent flow behavior.


\subsection{Sensitivity analysis} \label{ssec:sensitivity_analysis}

\begin{figure}[H]
\centering
\begin{subfigure}[b]{0.49\textwidth}
 \centering
 \includegraphics[width=\textwidth]{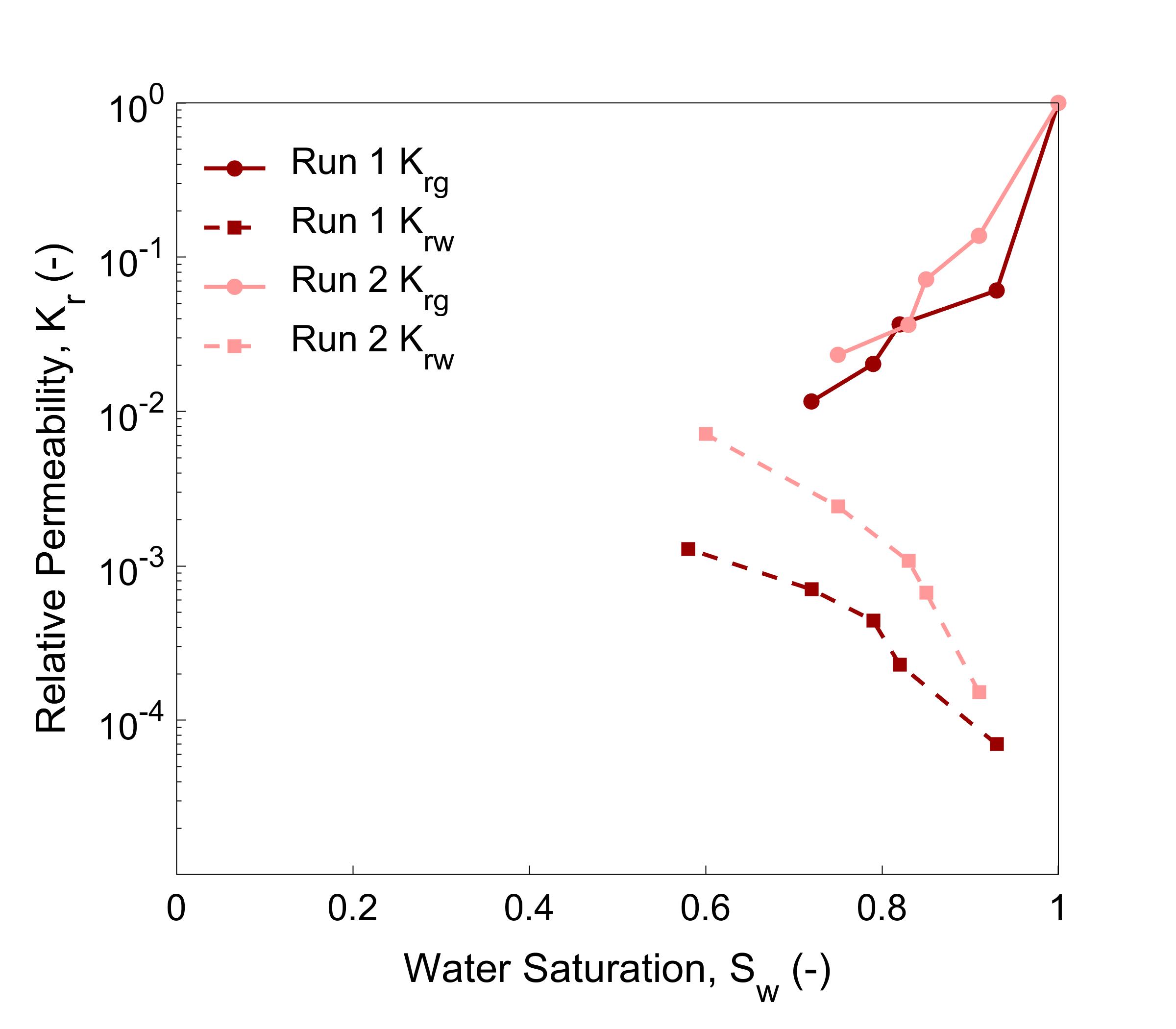} 
   \caption{}
\label{Relperm_sensitivity}
\end{subfigure}
\hfill 
\begin{subfigure}{0.49\textwidth}
 \centering
 \includegraphics[width=\textwidth]{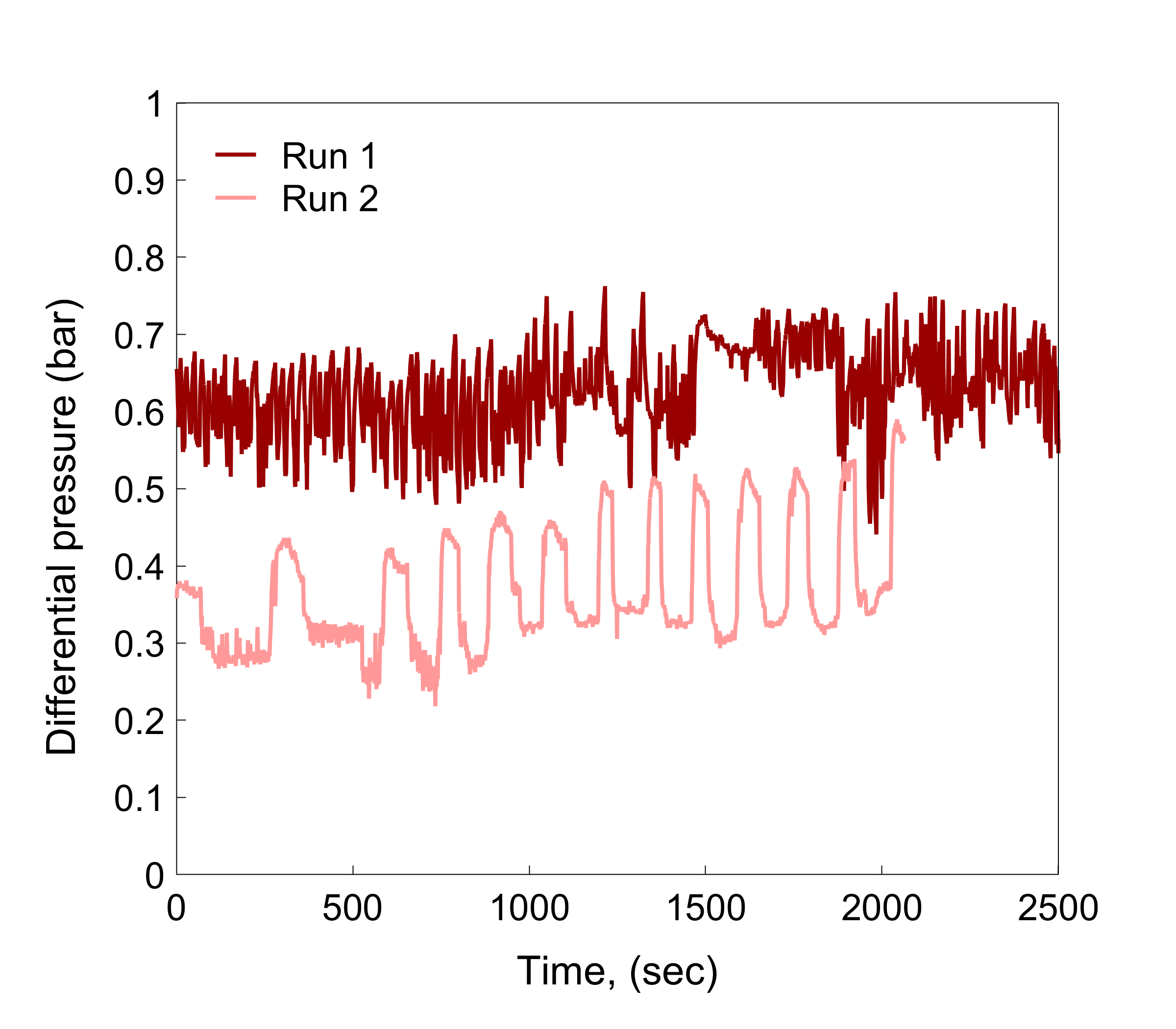} 
  \caption{}
\label{RelPerm_press_sensitivity_fg06}
\end{subfigure}
\caption{ (a) Relative permeability of H$_2$-brine system for two runs.  (b) Differential pressure recorded over 30 minutes for two runs of H$_2$-brine experiment.}
\label{RelPerm_sensitivity}
\end{figure}

As discussed in previous sections, relative permeabilities at reservoir conditions in fractures are strongly influenced by intermittency, leading to significant fluctuations in pressure and saturation. To assess the sensitivity of the hydrogen-brine system under these conditions, we repeated the experiment under identical flow conditions. Overall, qualitative trends in the relative permeability-saturation curves remained consistent across both runs, as shown in Figure \ref{RelPerm_sensitivity}. The water permeabilities ($K_{rw}$) were quantitatively similar in both runs. Gas permeability, while consistently low across runs, exhibited slight differences, likely due to minor pressure fluctuation variations. Since gas permeabilities are very low, even slight variations in pressure fluctuations can result in differences in the measured relative gas permeability. The pressure fluctuations for these two runs at $f_g = 0.6$ are shown in Figure \ref{RelPerm_press_sensitivity_fg06}. The data clearly exhibits cyclic fluctuations, with the timescales of these fluctuations differing between the two runs: in the first run, the period was 0.5 minutes, while in the repeated run, the period extended to 2.5 minutes. These variations in fluctuation timescales may contribute to slight differences in the calculated relative gas permeabilities. Such differences could arise due to slight changes in fracture morphology, potentially occurring during pressurization, as the fracture is sensitive to applied stress. Over the course of running different experiments presented in this paper , the rock sample underwent pressurization-depressurization cycles, potentially impacting fracture morphology. This is reflected in the observed decrease in absolute permeability before the experimental runs, from 23 D to 8 D. Despite these slight variations, the overall behavior remains consistent across both experimental runs, confirming that our workflow is robust and validating the results presented in this study. The 1-D saturation profiles for these two runs also confirm it as shown in Supplementary Information S3.

While ideal statistical validation would involve multiple repetitions for each gas - brine pair, such replication was not feasible due to the significant safety and logistical challenges of working with hydrogen at 10 MPa. Nevertheless, the observed intermittency and capillary trapping behavior were reproducible across two hydrogen experiments and consistent with trends observed in methane, a gas with comparable viscosity and interfacial tension. In contrast, nitrogen, which is supercritical and an order of magnitude higher viscosity exhibited significantly higher relative permeability. These results are also consistent with prior literature as further discussed in Section \ref{sec:discussion}.

\section{Discussion} \label{sec:discussion}

\begin{figure}[H]
 \centering
 \noindent\includegraphics[width=\textwidth]{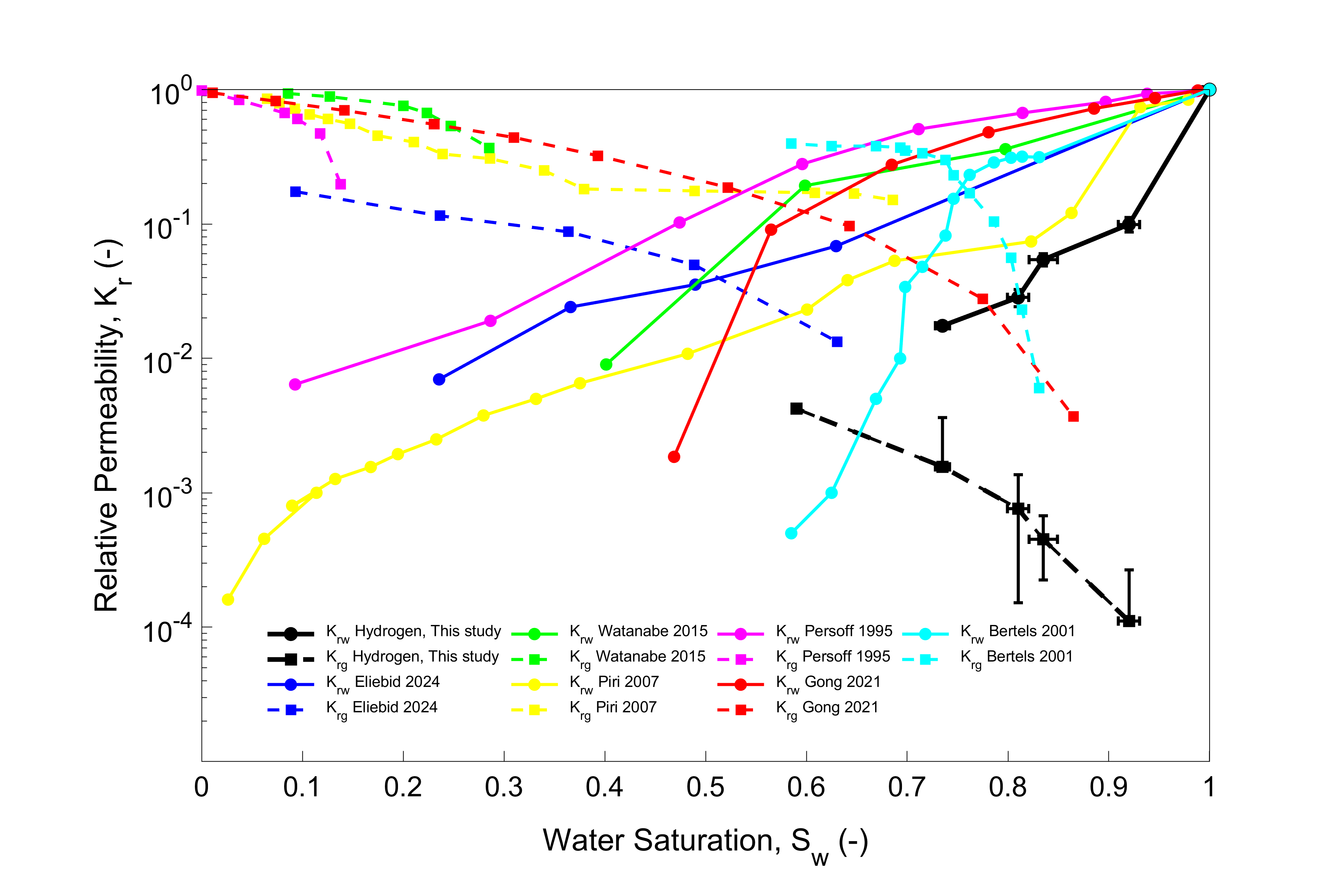}
  \caption{Comparison of relative permeability measurements with previous studies in fracture.}
\label{RelPerm_Fracture}
 \end{figure}

Direct measurements of relative permeability for hydrogen compared to other gases in fractured rock samples are currently absent from the literature. Here, we compare our findings with existing experimental data and numerical models for different fluid systems in rough fractures, as shown in Figure \ref{RelPerm_Fracture}. Compared to previous studies, the hydrogen gas relative permeability in our system is notably lower, indicating interference effects that are stronger than predicted by existing models such as \cite{Watanabe2015}, \citet{Gong2021}, and \citet{Bertels2001}. The limited available experimental data were collected under different flow conditions and using different fluids, making direct comparisons challenging. The fluid pairs used in various studies from the literature are summarized in Supplementary Information S4. The nitrogen permeability aligns more closely with the model proposed by \citet{Gong2021}; however, the impact of fracture geometry could not be fully resolved based on these comparisons. The nitrogen-brine system relative permeabilities were also compared to models from the literature, with detailed discussions provided in the Supplementary Information S4.  In contrast, water permeability in our results aligns more closely with the model suggested by \citet{Piri2007}. Importantly, fracture morphology also influences these differences, so while such comparisons are informative, they are not definitive. Controlled experiments and models that account for specific fracture morphology in hydrogen-brine systems are essential for more conclusive insights.

Notably, recent studies in porous media report endpoint relative permeabilities for hydrogen as low as 10$^{-3}$ \cite{Boon2022, Lysyy2022}, which aligns with our fracture flow observations and highlights hydrogen’s reduced connectivity compared to methane and nitrogen. For instance, \citet{Boon2022} reported H$_2$ - brine relative permeability in Berea sandstone to be an order of magnitude lower than CO$_2$ - water and N$_2$ - water systems in the same rock, as shown by \citet{Pini2013}. This indicates stronger phase interference for low viscosity gases such as hydrogen. Our measured fracture relative permeabilities, while lower than values reported in other fracture studies, are therefore consistent with trends observed in hydrogen-specific porous media literature. Thus, we further compare our results with the hydrogen specific literature of porous media supporting the interpretation of the specificity of hydrogen flow compared to other gases investigated in the literature.

\begin{figure}[H]
 \centering
 \noindent\includegraphics[width=\textwidth]{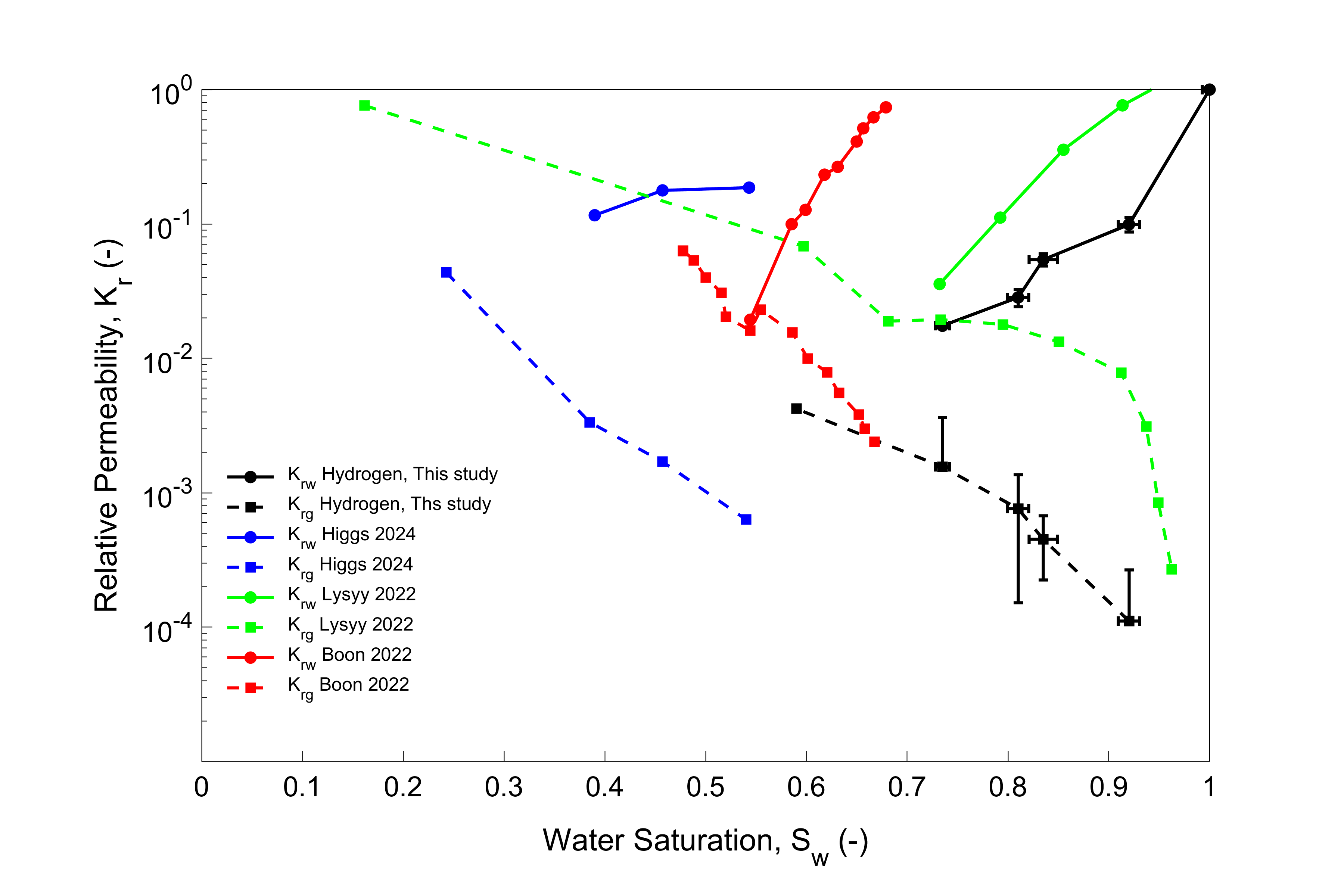}
  \caption{Comparison of relative permeability measurements with previous hydrogen measurement studies in porous media. The black lines denote the hydrogen-brine data from our study.}
\label{RelPerm_Porous}
 \end{figure}

The lower hydrogen gas permeability observed in our fracture study is consistent with trends seen in porous media studies in Berea sandstone \cite{Lysyy2022, Boon2022} and Bentheimer sandstone \cite{Higgs2023}, shown in Figure \ref{RelPerm_Porous}. Despite marked differences in sample geometry and flow conditions, the results remain qualitatively comparable, as all experiments were conducted at low capillary numbers (Ca$_t$). 
In porous media, the relative permeability curves from \citet{Higgs2023} exhibit closer alignment for hydrogen, methane, and nitrogen. However, our fracture study reveals greater differences in both wetting- and non-wetting-phase permeabilities. This divergence can likely be attributed to the increased intermittency in the fracture, resulting from the higher pressure conditions (10 MPa) in our study, where the interfacial tension (IFT) between hydrogen and methane differs more significantly, and nitrogen exists in a supercritical state. In contrast, the experiments by \citet{Higgs2023}, conducted at 20 bar, were performed under conditions where the fluid properties of these gases were more similar. Additionally, our study shows a much higher irreducible saturation, exemplifying the dependence of capillary-dominated relative permeability curves on flow rate (injection rate) and geometry (porous vs. fractured). When comparing our hydrogen permeability results to those from \citet{Boon2022}, who also conducted high-pressure experiments (10 MPa) at 18$^{\circ}$C under similar conditions, hydrogen and water permeabilities were generally higher across most saturations for porous media, though the general trend remains consistent. These comparisons highlight the increased complexity of fluid interactions in fractures compared to those observed in porous media.

Low hydrogen gas relative permeability ($K_{rg}$) directly limits its mobility, requiring higher injection pressures to meet target storage rates. Studies in porous media, including CO$_2$ storage simulations \cite{Oostrom2016, Boon2021} and recent hydrogen-specific modeling \cite{Bo2023}, show that low endpoint $K{rg}$ values lead to compact, flatter gas plumes with relatively uniform saturation and low gas saturations beyond the dry-out zone. While this reduces storage efficiency, it also spatially confines the gas plume, potentially lowering leakage risks and simplifying monitoring. \citet{Bo2023}, in a porous reservoir context, further demonstrated that increasing injection or production rates does not necessarily enhance performance due to limited flow and pressure distribution caused by gas trapping. Using relative permeability data from other fluid systems can therefore introduce significant errors in predicting hydrogen behavior. Moreover, compensating for low mobility by injecting at higher rates could risk exceeding caprock integrity thresholds. These insights highlight the need to extend such simulation studies to fractured reservoirs, using laboratory-measured hydrogen relative permeability data under relevant flow conditions. Accurate characterization of $K_{rg}$ in site-specific fracture systems is critical to avoid underestimating injectivity limitations or overestimating storage capacity, and to ensure a realistic balance between operational feasibility and storage efficiency.

\section{Conclusions} \label{sec:conclusions}

The findings from our hydrogen-brine and methane-brine experiments, along with nitrogen-brine experiment, provide crucial insights for hydrogen storage in subsurface fractured reservoirs and particularly in the case of transitioning existing natural gas storage facilities. Our experiments are focused on the Loenhout storage site in northern Belgium which is a part of the BeHyStore project. We conducted direct measurements of hydrogen, methane, and nitrogen relative permeabilities in fractured rock under reservoir pressure conditions (10 MPa), while recognizing that experimental limitations prevented matching the reservoir temperature. However, the impact of temperature on relative permeability is expected to be minimal.

Our study summarizes three key findings relevant to hydrogen storage in fractured reservoirs:

\begin{enumerate}

\item{The fracture relative permeability curves for hydrogen and methane are similar, suggesting that similar intrinsic flow properties would be maintained after repurposing the storage from natural gas to hydrogen. The minor differences observed are attributed to methane’s higher viscosity and lower interfacial tension (0.05 N m$^{-1}$ for methane versus 0.07 N m$^{-1}$ for hydrogen).}
\item{Relative permeability for both hydrogen and methane in fractured rocks is low, with substantial phase interference caused by flow intermittency. This behavior is not well represented by existing models, highlighting the need for more advanced modeling approaches to accurately capture the dynamics of hydrogen and methane in fractured media.}
\item{Nitrogen exhibits significantly higher gas relative permeability, making it unsuitable as a laboratory proxy. It also influences considerations for its use as a  cushion gas due to its propensity for breakthrough issues during injection and production.}

\end{enumerate}

In underground gas storage, tight shale caprock typically acts as a barrier, but fractures, whether preexisting or induced by pressure variations during hydrogen injection, can provide additional pathways for gas movement. The low gas relative permeability observed in this study may contribute to the mitigation of leakage risks supporting the long-term stability and security of hydrogen storage.

While this study utilizes a single artificially fractured sample, which cannot capture the full spectrum of natural fracture heterogeneity (e.g., aperture distributions, mineral heterogeneity, or network connectivity), its primary objective was to compare fluid-specific behavior of hydrogen, methane, and nitrogen under as identical fracture conditions as possible. The observed differences in relative permeability, intermittency, and endpoint saturation are therefore attributable to intrinsic gas properties (e.g., viscosity, dissolution, interfacial tension) rather than geometric variability. Importantly, the sample features a realistic branching network (Figure \ref{Dry3DScan}) and significant roughness, providing important insights into gas trapping and connectivity. Given the absence of high-pressure hydrogen relative permeability data for fractures, our work provides a foundational dataset to validate pore-scale hydrogen transport models, support discrete fracture network (DFN) simulations that upscale permeability based on aperture distributions \cite{Hyman2015}, and guide future experimental design across varied fracture morphologies.

In conclusion, this study provides critical insights into hydrogen behavior in fractured rock, supporting efforts to optimize underground hydrogen storage. By highlighting the limitations of using nitrogen as a proxy for hydrogen, our findings emphasize the need for hydrogen-specific research to develop effective storage strategies and safety protocols in subsurface reservoirs. In this study, we considered only drainage relative permeabilities. However, cyclic operations are likely to induce hysteresis, particularly in the gas phase, as observed by \citet{Boon2024} in porous sandstone. A comprehensive understanding of cyclic underground hydrogen storage (UHS) at field scale would therefore require future work to measure or simulate imbibition relative permeabilities. In addition to imbibition relative permeabilities, the future work should focus on measuring hydrogen-brine relative permeability in fracture samples with varied aperture sizes and roughness. Such studies will be valuable in understanding the impact of fracture geometry on flow behavior, ultimately enhancing the accuracy of upscaled fracture network models for reservoir-scale applications.

  
\section*{Acknowledgments}
This study received funding from the Research Foundation-Flanders (FWO projects G004820N), the Energy Transition Fund of the Belgian Federal Governement (project BE-HyStore) and Ghent University's special Research Fund (BOF/COR/2022/008). SB acknowledges support from the FWO under fellowship 1S30825N. HO and TB acknowledge support from the FWO under project G037222N. TB acknowledges financial support from Fluxys, Belgium. We also acknowledge Prof. Kamaljit Singh and Zaid Jangda (Heriot-Watt University) for their guidance in designing the hydrogen flow setup and Yves Israel (Department of Physics and Astronomy, Ghent University) for technical assistance with the experimental setup.

\bibliographystyle{unsrtnat}

\bibliography{manuscript_revisions}


\end{document}